# Nitrogen-Triggered Amorphization Enables High-Performance Solid-State Electrolytes

Bolong Hong[1,2], Lei Gao[3]*, Bingkai Zhang[4], Pengfei Nan[5], Ruishan Zhang[4], Yuhang Li[6], Zhihao Lei[6], Ming Liu[6], Jing Wu[7], Longbang Di[3], Haijin Ni[2], Songbai Han[2]*, Jinlong Zhu[2,8]*

[1]*College of Semiconductors (National Graduate College for Engineers), Southern University of Science and Technology, Shenzhen 518055, China.*

[2]*Shenzhen Key Laboratory of Solid State Batteries, Guangdong Provincial Key Laboratory of Energy Materials for Electric Power, Guangdong-Hong Kong-Macao Joint Laboratory for Photonic Thermal-Electrical Energy Materials and Devices, Southern University of Science and Technology, Shenzhen 518055, China.*

[3]*School of Advanced Materials, Peking University, Shenzhen Graduate School, Shenzhen 518055, China.*

[4]*School of Chemical Engineering and Light Industry, Guangdong University of Technology, Guangzhou, 510006, China.*

[5]*Information Materials and Intelligent Sensing Laboratory of Anhui Province, Leibniz International Joint Research Center of Materials Sciences of Anhui Province, Institutes of Physical Science and Information Technology, Anhui University, Hefei 230601, China.*

[6]*Shenzhen Geim Graphene Center, Tsinghua University, Shenzhen International Graduate School, Shenzhen 518055, China.*

[7]*Cryo-EM Center, Southern University of Science and Technology, Shenzhen, Guangdong, 518055, China.*

[8]*Department of Physics, Southern University of Science and Technology, Shenzhen 518055, China.*

*Corresponding author

Correspondence to: Lei Gao, Songbai Han, Jinlong Zhu

gaolei2018@pku.edu.cn (L. Gao); hansb@sustech.edu.cn (S. Han); zhujl@sustech.edu.cn (J. Zhu)

## Abstract

Amorphous solid-state electrolytes (SSEs) hold great promise for advancing the application of all-solid-state batteries (ASSBs), owing to their favorable ionic conductivity, structural tunability, and promising electrochemical performance. However, the absence of universal design principles for amorphous SSEs limits their development. By fundamentally re-evaluating the amorphization-forming ability of amorphous SSE systems, this study establishes a nitrogen-driven universal strategy to convert diverse metal chlorides into amorphous $x$$Li_3N$-$MCl_y$ ($0.3 \leq 3x \leq 1.9$; M denotes a metal element; $2 \leq y \leq 5$) SSE. Nitrogen synergistically disrupts crystalline order via distorted coordination polyhedra and N-bridged networks, while dynamic bond reorganization enables rapid $Li^+$ migration, achieving ionic conductivity of 2.02 mS $cm^{-1}$ for 0.533$Li_3N$-$HfCl_4$ at 25 °C. Structural-property relationships reveal that high charge density and bridging capability of $N^{3-}$ enhance network disorder, shorten metal-coordinating atom distances, and optimize $Li^+$ diffusion pathway connectivity. ASSBs employing 0.533$Li_3N$-$HfCl_4$ retain 81.87% capacity after 2000 cycles at 1,000 mA $g^{-1}$ with high cathode loading (6.24 mg $cm^{-2}$), demonstrating engineering viability. This work provides a paradigm for rational design of high-performance amorphous SSEs.

## Introduction

The pursuit of high-performance solid-state electrolytes (SSEs) is a cornerstone in developing next-generation all-solid-state batteries (ASSBs) featuring high safety and energy density[1,2]. In the past decade, inorganic crystalline SSEs have attracted significant research attention owing to their promising properties. However, intrinsic limitations, including grain boundary resistance, mechanical deformability, interface compatibility, and scalable synthesis continue to hinder their applications[3-5]. For instance, crystalline oxide-based SSEs (e.g., garnet and NASICON-type) are limited by

high synthesis temperatures, high grain boundary impedance, and poor wettability by Li metal[6]. Although sulfide-based SSEs (e.g., $Li_{10}GeP_2S_{12}$, $Li_6PS_5Cl$) present ionic conductivities comparable to those of liquid electrolytes, the narrow electrochemical stability windows and poor stability against moisture in the air limit their applications in ASSBs[7].

Recent advances in amorphous SSEs offer a promising alternative to overcome the limitations hindering the development of ASSBs[7,8]. Specifically, amorphous SSEs feature isotropic ion transport without grain boundary constraints[9,10], leading to higher conductivity and lower temperature sensitivity[11]. In terms of interfacial compatibility, their low interfacial heterogeneity favors compatibility with electrodes[12]. Moreover, the absence of grain boundary defects outstand itself as SSE with higher density, improved mechanical strength, and better processability[13].

Still, fundamental limitations persist across a wide range of amorphous SSE systems, despite recent notable progress in their development. Sulfide-based amorphous SSE (e.g., $Li_2S$-$P_2S_5$ glasses) achieve exceptional ionic conductivities (> 1 mS cm$^{-1}$ at 25 °C) via disordered thiophosphate networks, but their narrow electrochemical stability (<3.5 V vs. $Li^+$/Li) hinders integration with high-voltage cathodes[14]. Oxide amorphous systems (e.g., $Li_3BO_3$-$Li_2SO_4$ glasses) exhibit enhanced oxidative stability (> 4.0 V) and mechanical robustness[15], yet their ionic conductivities remain below 0.1 mS cm$^{-1}$ at room temperature (RT). Recently, a new class of amorphous SSEs, including Cl-based and dual-anion-based systems (labeled to $Li_xA$-$MCl_y$, where $x$ = 1, 2, 3; A = Cl, O, N; M denotes a metal element; $y$ = 2, 3, 4, 5), has attracted considerable interest due to their structural diversity and promising electrochemical properties (Fig. 1)[7,16]. These amorphous electrolytes are typically synthesized using lithium-containing precursors and metal chlorides as starting materials through high energy ball milling. During

synthesis, ion exchange occurs between lithium-containing precursors and metal chlorides at specific molar ratios, resulting in the distortion of coordination polyhedra and a formation of a disordered network. This configuration generates an amorphous structure with fast $Li^+$ diffusion channels, leading to enhanced performance compared with crystalline chloride-based SSEs[17]. For instance, O-Cl based SSEs systems synthesized by introducing $Li_2O$ into metal chlorides (e.g., Li-Zr-O-Cl[13,18], Li-Ta-O-Cl[19]) demonstrate breakthrough ionic conductivities (> 1 mS $cm^{-1}$ at 25 °C), high-voltage stability, and deformability. However, such strategies remain limited to a narrow subset of metal chlorides (e.g., $TaCl_5$, $NbCl_5$, $ZrCl_4$, $HfCl_4$)[12,19,20]. In addition, our recent work identified a new amorphous N-Cl based SSE, in which $0.417Li_3N$-$TaCl_5$ exhibited high ionic conductivity of 7.34 mS $cm^{-1}$ at 30 °C and reserved performance at low temperature[11,21]. Critically, although new amorphous SSEs continue to be reported, the synthesis and formation principles of amorphous halides and dual-anion based SSEs, as well as the specific mechanism of $Li^+$ migration remain to be fully established[22]. A key challenge is the lack of a universal methodology for systematically synthesizing amorphous SSEs and establishing a systematic framework for the design of amorphous SSEs. Large-scale systematic modeling is also essential for unveiling the mechanisms underlying the formation of amorphous SSEs and providing profound insights into $Li^+$ transport dynamics. This understanding will be critical for a rational design and synthesis of next-generation high-performance amorphous SSEs.

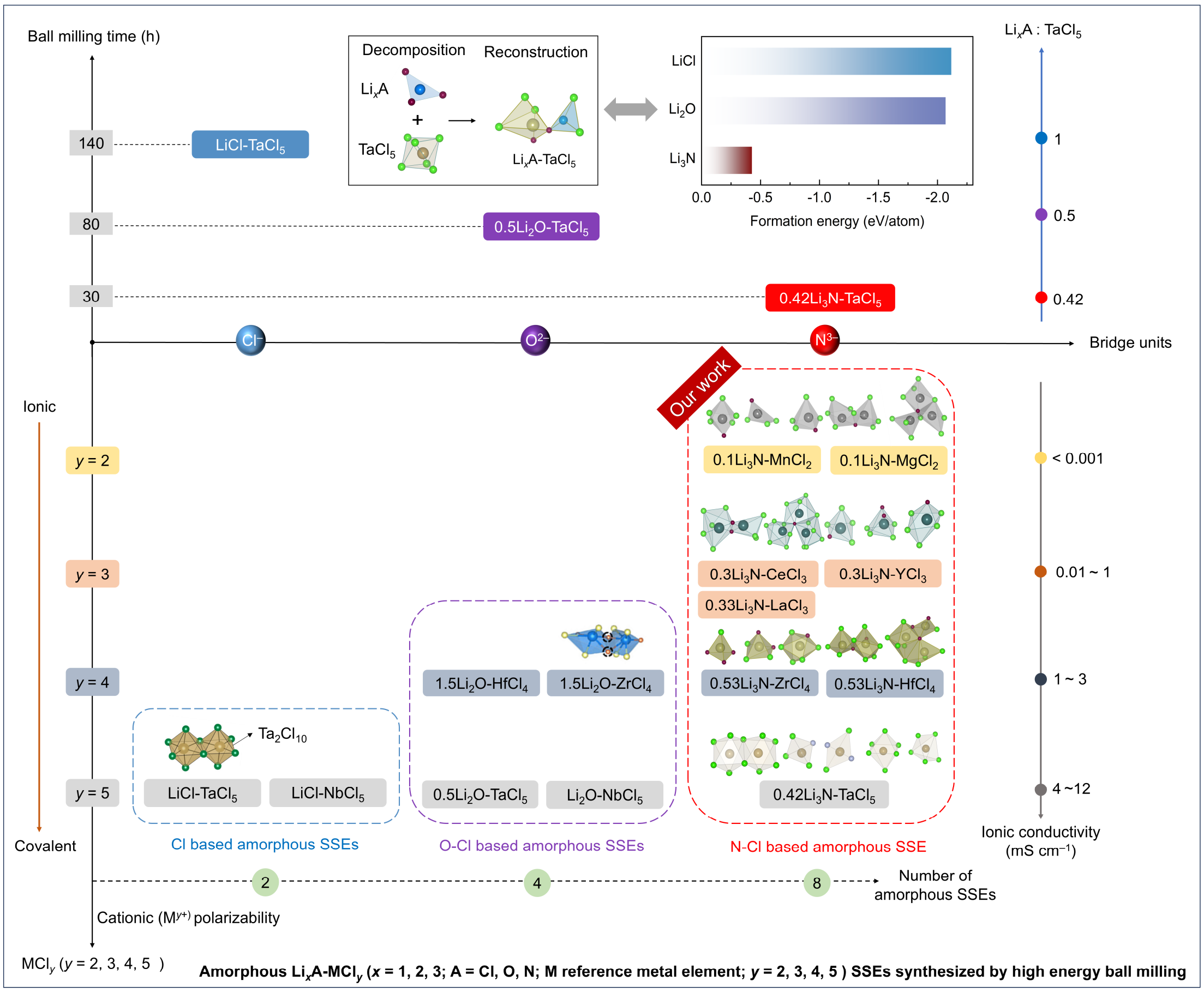


**Figure 1. Roadmap for the recent advancement of Amorphous $Li_xA$-$MCl_y$ ($x$ = 1, 2, 3; A = Cl, O, N; M denotes a metal element; $y$ = 2, 3, 4, 5) SSEs synthesized by high energy ball milling.** SSE compositions and conductivity data are from this work and prior literature[18,20,21,23]. Ball milling time data for 0.5$Li_2O$-$TaCl_5$ are from Supplementary Fig. 1, while other data are from literature[11]. The formation energy data were obtained from the Materials Project database[24].

Here, we unveil an amorphization mechanism in $Li_xA$-$MCl_y$ SSEs driven by covalent interactions and bridging structures (Fig. 1), and establish a universal N-driven strategy for the formation of amorphous $x$$Li_3N$-$MCl_y$ SSEs ($0.3 \leq 3x \leq 1.9$; M denotes a metal element; $2 \leq y \leq 4$). Benefiting from the high amorphization-forming ability (AFA) of the $x$$Li_3N$-$MCl_y$, $N^{3-}$ bridged networks induce structural disorder and enable fast $Li^+$ migration through dynamic bond reorganization, achieving ionic conductivities up to

2.02 mS cm$^{-1}$ (0.533$Li_3N$-$HfCl_4$). Based on the ionic conductivity trends of a series of amorphous $x$$Li_3N$-$MCl_y$ SSEs, the structure-property relationship of amorphous SSEs is comprehensively investigated. In addition, ASSBs with the configuration of Li-In|$Li_{5.5}PS_{4.5}Cl_{1.5}$|0.533$Li_3N$-$HfCl_4$|$LiNi_{0.83}Mn_{0.05}Co_{0.12}O_2$ employing high-loading cathodes (6.24 mg cm$^{-2}$), exhibit a capacity retention of 81.87% after 2,000 cycles at a high current density of 1,000 mA g$^{-1}$. This approach of nitrogen-driven amorphization defines a universal design principle for high-performance amorphous SSEs.

## Results and discussion

### Role of network modifiers and formers in facilitating amorphization

High-energy ball milling is currently the primary method for preparing amorphous SSEs, leveraging localized mechanical stress and thermal effects to drive particle comminution and solid-phase reactions[25,26] (Supplementary Fig. 2). However, the complex interplay of stress, heat, and atomic interactions within the milling process makes it challenging to observe and investigate the mechanisms of amorphous phase formation in real time. Despite these challenges, well-defined starting materials and synthesis conditions enable systematic studies of amorphization mechanisms through comparative analysis. Theoretical frameworks, notably the continuous random network (CRN) model proposed by Zachariasen[27], elucidate the structural principles governing amorphous phase formation, emphasizing the necessity of a continuous disordered atomic network for glass structures. This model highlights the critical role of AFA or glass-forming ability, which is defined as the ease with which a material system transforms to an amorphous state[28]. This model highlights the critical role of AFA or glass-forming ability, which depends on the contributions of amorphous network formers and modifiers in stabilizing the disordered structure[29].

AFA plays a vital role in the design and synthesis of high-performance amorphous

SSEs. However, the understanding of AFA in the formation of amorphous $Li_xA$-$MCl_y$ systems remains insufficiently explored. During ball milling synthesis of $Li_xA$-$MCl_y$ SSEs, metal chlorides dissociate lithium-containing precursors and release free $Li^+$ ions[30]. Concurrently, anions (from $Li_xA$) are incorporated into the metal chloride, forming coordination polyhedra, such as $[TaCl_6]^-$ in LiCl-$TaCl_5$[20,31]. Within the resultant amorphous network, $Li^+$ ions function as network modifiers that disrupt lattice symmetry and confer ionic conductivity to the structure, while these coordination polyhedra act as network formers that establish the continuous and disordered framework[32]. As fundamental building units of the disordered network, the bonding nature of the network formers is a core factor influencing AFA. When comparing covalent and ionic bonds, which are commonly present in SSEs, structures dominated by covalent bonds exhibit significantly stronger AFA[33]. This is primarily due to the tendency of covalent compounds to form three-dimensional network structures, such as $SiO_2$ glass. In such a system, the Si–O bonds prefer corner-sharing tetrahedral configurations, and the random distortions in Si–O–Si bond angles promote structural disorder[34]. Furthermore, covalent bonds readily form long chains or extended networks (e.g., polymers), whose inherent folding and bending characteristics easily induce long-range disorder, which explains the strong AFA typically observed in polymers[35]. Fig. 1 indicates that Cl-based amorphous LiCl-$MCl_y$ SSEs are rare, and tend to form only in the presence of covalent pentavalent metal chlorides[36], such as $TaCl_5$, which act as starting material. The covalent Ta–Cl bonds provide a critical bonding framework for the formation of the network former $[TaCl_6]^-$. In contrast, the tetravalent metal chloride $HfCl_4$ exhibits more ionic character, attributed to the lower electronegativity of Hf. Consequently, the $x$LiCl-$HfCl_4$ system possesses weaker AFA, leading to the formation of crystalline $Li_2HfCl_6$[37]. Similarly, ionic crystals such as $YCl_3$ and $MgCl_2$ resist amorphization upon milling with LiCl[38,39], as their highly ordered ionic coordination creates stable crystalline structures that hinder the long-range disorder required for amorphous phases[33].

Moreover, introducing high charge density bridging anions (e.g., $O^{2-}$) into metal chlorides offers a strategy to enhance the AFA of the SSE system. These anions distort metal-centered coordination polyhedra and bridge multiple polyhedral units, thereby generating diverse network formers (e.g., $[TaCl_{5-a}O_a]^{a-}$; $1 \leq a < 5$) that stabilize amorphous structures[19]. In particular, the number of O-Cl dual-anion amorphous SSEs is twice as abundant as chloride-based SSEs (Fig. 1). The formation of amorphous $x Li_2O$-$HfCl_4$/$ZrCl_4$ SSEs relies on the synergistic effect between the residual covalency in Hf/Zr-Cl bonds and the oxygen bridging, which extends to tetravalent chlorides but not ionic trivalent chlorides, such as $YCl_3$ and $LaCl_3$. For instance, ball milling $Li_2O$ with $YCl_3$ forms crystalline $Li_3YCl_3O_{1.5}$ with low $Li^+$ conductivity[18]. The AFA imparted by a network former is governed by the charge density of its bridging anion. Bridging anions with higher charge density for bonding can concurrently coordinate more metal centers and generate more complex network formers, thereby imparting stronger AFA to the system while requiring fewer such bridging anions. Consequently, $O^{2-}$ does not have sufficient bridging capability to induce amorphization in ionic divalent/trivalent metal chlorides. Interestingly, our prior work demonstrated that $N^{3-}$ (in $Li_3N$), possessing higher charge density, readily forms amorphous $0.33Li_3N$-$LaCl_3$ SSEs when milled with ionic $LaCl_3$[21], indicating that the $x Li_3N$-$MCl_y$ system has strong AFA. Furthermore, when $TaCl_5$ and lithium-containing precursors are used as the starting materials, the $Li_3N$ can significantly improve the efficiency of amorphous SSEs compared with LiCl or $Li_2O$ (Fig. 1)[11]. This is due to the higher formation energy of $Li_3N$, which allows for easier decomposition and efficient ion exchange with $TaCl_5$, with the assistance of high energy ball milling to across its kinetic energy barrier, thus accelerating the formation of the amorphous phase[20].

Taken together, both theoretical analysis and experimental results demonstrate the dual advantages of $x$$Li_3N$-$MCl_y$ systems: its superior AFA, driven by the high charge density $N^{3-}$ bridging, and its facile synthesis, enabled by high formation energy of $Li_3N$ (Fig. 1). These attributes suggest the potential for developing a series of novel N-Cl amorphous SSEs through the incorporation of $Li_3N$ into various metal chlorides, including ionic chlorides that are typically challenging to be amorphized. This N-driven synthesis approach offers a valuable material platform for conducting a systematic and in-depth study of the relationship between the performance and composition of amorphous SSEs.

**Nitrogen-driven synthesis of amorphous solid-state electrolytes**

To investigate the AFA of $x$$Li_3N$-$MCl_y$ system, we selected $Li_3N$ and a range of metal chlorides with varying ionic characteristics (e.g., $HfCl_4$, $CeCl_3$, $MgCl_2$), to serve as starting materials through the high energy ball milling synthesis of amorphous $x$$Li_3N$-$MCl_y$ SSEs. Unlike other amorphous SSEs, formed via extended ball milling, such as those in LiCl-$TaCl_5$ systems[20,23] or hazardous gas-generating reactions in LiOH-$ACl_5$ (A = Ta, Nb)[40] or $MAlCl_{4-2x}O_x$ (M = Li, Na)[41] routes, the N-driven protocol efficiently converts diverse metal chlorides into phase-pure amorphous SSEs in this work. Various metal chlorides were successfully demonstrated, including $TaCl_5$, $HfCl_4$, $ZrCl_4$, $CeCl_3$, $LaCl_3$, $YCl_3$, $MgCl_2$, and $MnCl_2$ (Fig. 1), with over 30 theoretically compatible candidates (Supplementary Fig. 3). Remarkably, pentavalent metal chlorides (e.g., $TaCl_5$) exhibit intrinsic amorphization tendencies with diverse lithium-containing precursors due to their high covalent character (Fig. 1)[23,42]. However, the inherent characteristic obscures the individual roles of the bridging function and the covalency

in AFA. Based on this fact, to more clearly investigate the attribution of the amorphous phase formation to N-mediated coordination engineering, we focus on $MCl_y$ (M = metal elements, $2 \leq y \leq 4$) systems with inherently ionic character.

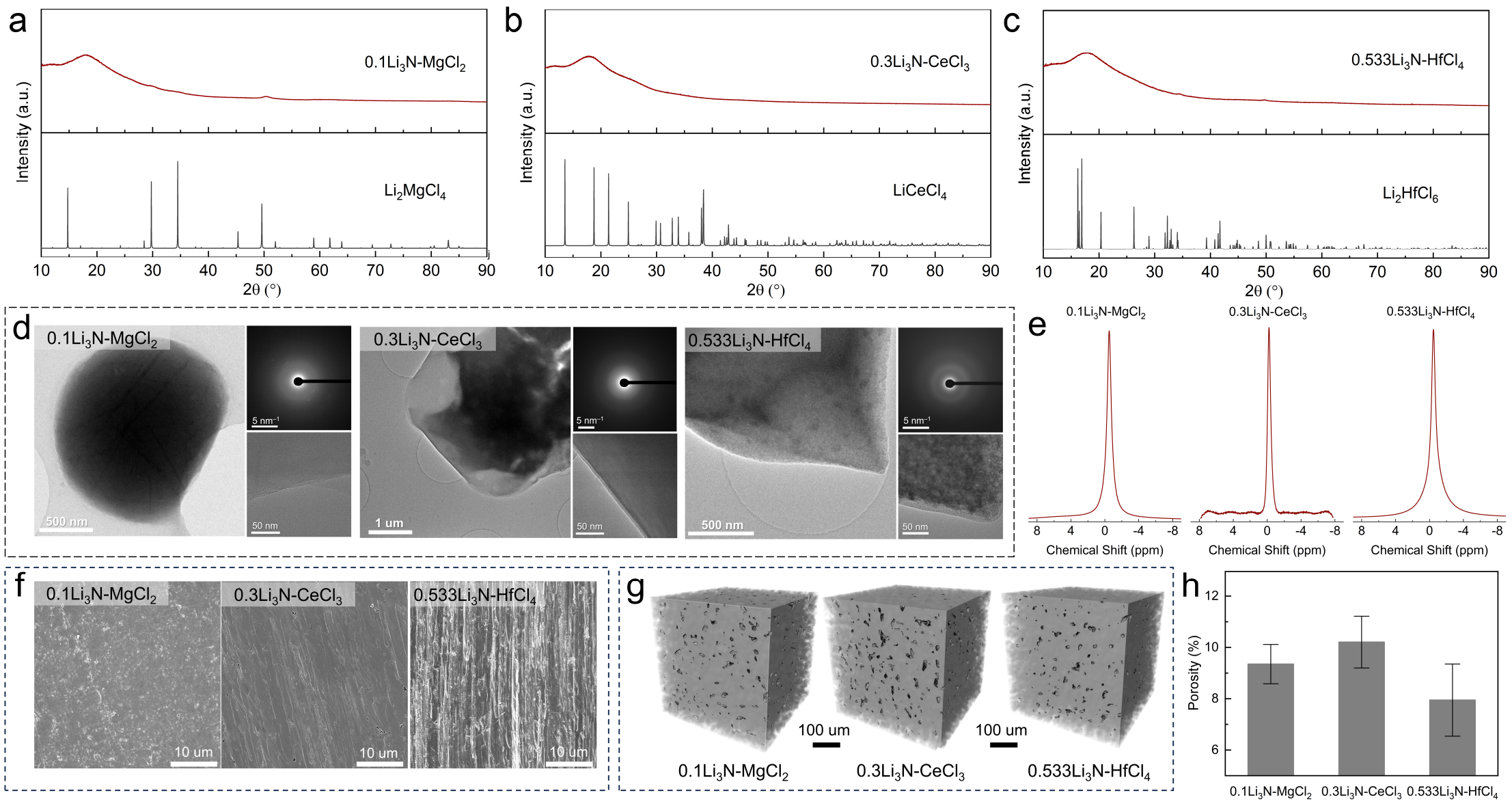

**Figure 2. Synthesis and structure analysis of amorphous $x$$Li_3N$-$MCl_y$ SSEs.** (a-c) XRD patterns of representative 0.1$Li_3N$-$MgCl_2$ and $Li_2MgCl_4$ (a), 0.3$Li_3N$-$CeCl_3$ and $LiCeCl_4$ (b), 0.533$Li_3N$-$HfCl_4$ and $Li_2HfCl_6$ (c). The structural data for the crystalline SSEs were obtained from the ICSD and OQMD databases[43,44]. (d) Cryo-TEM images and corresponding FFT patterns of amorphous 0.1$Li_3N$-$MgCl_2$, 0.3$Li_3N$-$CeCl_3$, and 0.533$Li_3N$-$HfCl_4$. (e) $^7Li$ solid-state NMR spectra of the amorphous 0.1$Li_3N$-$MgCl_2$, 0.3$Li_3N$-$CeCl_3$, and 0.533$Li_3N$-$HfCl_4$. (f) SEM images of cold-pressed $x$$Li_3N$-$MCl_y$ SSEs pellets. (g) 3D volume-rendered images from XCT scan. (h) Porosity analyzed by XCT for 0.1$Li_3N$-$MgCl_2$, 0.3$Li_3N$-$CeCl_3$, and 0.533$Li_3N$-$HfCl_4$.

XRD patterns of $x$$Li_3N$-$MCl_y$ ($0.3 \leq 3x \leq 1.9$; M = Hf, Ce, Mg; $2 \leq y \leq 4$) in Supplementary Fig. 4, plus representative 0.1$Li_3N$-$MgCl_2$, 0.3$Li_3N$-$CeCl_3$, and 0.533$Li_3N$-$HfCl_4$ in Fig. 2a-c, all exhibit completely amorphous features, contrasting with the crystalline SSEs ($Li_2MgCl_4$, $LiCeCl_4$, $Li_2HfCl_6$) that display sharp diffraction features. Also, this amorphous feature is observed in other N-Cl based SSEs—including $0.1Li_3N-MnCl_2$, $0.3Li_3N-YCl_3$, and $0.533Li_3N-ZrCl_4$—as evidenced by the lack of sharp diffraction peaks in XRD profiles (Supplementary Fig. 5). Notably, the amorphous SSEs derived from $MgCl_2$, $YCl_3$, and $CeCl_3$ represent the first reported instances in their respective material classes, expanding the compositional landscape of amorphous halide SSEs (Fig. 1). Cryo-transmission electron microscopy (Cryo-TEM) further supports these findings, as $0.3Li_3N-CeCl_3$ and $0.533Li_3N-HfCl_4$ reveal disordered lattice structures and lacking crystalline fringes. The corresponding fast Fourier transform (FFT) patterns indicate the absence of diffraction spots (Fig. 2d). $0.1Li_3N-MgCl_2$ exhibits minimal residual crystallinity manifested as sporadic diffraction spots, likely attributable to the low formation energy of $MgCl_2$ (Supplementary Fig. 6). This thermodynamic stability limits its interaction with $Li_3N$, imposing kinetic barriers to complete amorphization[20]. The chemical homogeneity of these phases is further confirmed by $^7Li$ solid-state NMR spectra (Fig. 2e), which show a single Lorentzian peak, excluding residual lithium-containing impurities or structural inhomogeneities. Obviously, the AFA of the N-mediated strategy surpasses that of O- or Cl-based approaches, enabling the amorphization of a broader range of metal chlorides (Fig. 1 and Supplementary Fig. 3). In addition, Similar to other reported amorphous SSEs, $0.1Li_3N-MgCl_2$, $0.3Li_3N-CeCl_3$, and $0.533Li_3N-HfCl_4$ exhibit good compactness. Scanning electron microscopy (SEM) images of cold-pressed pellets at 500 MPa (Fig. 2f and Supplementary Fig. 7) reveal smooth surfaces with minimal void formation. Complementary X-ray computed tomography (CT) analysis (Fig. 2g and h) quantifies the three-dimensional porosity as 9.4% for $0.1Li_3N-MgCl_2$, 10.0% for $0.3Li_3N-CeCl_3$, and 7.9% for $0.533Li_3N-HfCl_4$, all notably lower than that of $Li_3InCl_6$ (~18%) and $Li_6PS_5Cl$ (~15%)[11].

**Decoding the role of nitrogen in coordination modification and bridging function**

To further elucidate the promoting role of nitrogen in amorphous phase formation, we conducted a comprehensive analysis of the structural characteristics and amorphization mechanisms of $x Li_3N$-$MCl_y$ SSEs. Figure 3a presents the Raman spectra of $x Li_3N$-$MCl_y$ SSEs alongside those of the corresponding crystalline metal chlorides. In crystalline $MgCl_2$, a sharp peak at 239.3 $cm^{-1}$ reflects well-defined metal-Cl lattice vibrations, arising from its long-range translational symmetry. Similarly, crystalline $CeCl_3$ exhibits distinct peaks at 101.1, 183.2, and 208.6 $cm^{-1}$, and $HfCl_4$ shows peaks at 79.0, 104.7, 121.2, 144.9, 284.3, and 389.7 $cm^{-1}$, all indicative of ordered metal-Cl vibrational modes characteristic of their crystalline lattices[12,45,46]. Upon $Li_3N$ incorporation, these sharp peaks broaden into diffuse bands, signaling a loss of long-range periodicity. This peak broadening reflects increased structural disorder, with a diversification of short-range coordination environments, bond angles, and lengths, resulting in a broader distribution of vibrational frequencies[47,48]. In amorphous $0.3Li_3N$-$CeCl_3$ and $0.533Li_3N$-$HfCl_4$, new broad features emerge at 300–500 $cm^{-1}$ and 400–650 $cm^{-1}$ (Supplementary Fig. 8), respectively, reflecting metal-N coordination formation[49,50], while $0.1Li_3N$-$MgCl_2$ lacks these signals, due to fluorescence interference obscuring the Mg-N vibrational signal.

Moreover, ab initio molecular dynamics (AIMD) simulations were employed to model the amorphous structure of $0.1Li_3N$-$MgCl_2$, $0.3Li_3N$-$CeCl_3$, and $0.533Li_3N$-$HfCl_4$ (Supplementary Fig. 9). Experimentally, pair distribution function (PDF) analysis, including partial PDF for M-Cl and M-N distances (M = Mg, Ce, Hf), was conducted to reveal the local coordination structure (Fig. 3b and c). In crystalline $Li_A MCl_B$ ($A$ = 1 or 2, M = Mg, Ce, Hf, $B$ = 4 or 6), the M-Cl pairs correspond to the nearest distances. In contrast, the M-Cl pairs in amorphous SSEs exhibit expanded ranges, reflecting N-induced distortion of coordination geometries and disruption of crystalline periodicity.

Specifically, the M-Cl distances range from 2.10 to 3.30 Å in 0.1$Li_3N$-$MgCl_2$, 2.42 to 3.67 Å in 0.3$Li_3N$-$CeCl_3$, and 2.17-3.46 Å in 0.533$Li_3N$-$HfCl_4$, respectively. PDF analysis further underscores the direct bonding interactions between nitrogen and metal cations (Fig. 3c), with nearest distances ranging from 1.9 to 2.32 Å for Mg-N, 2.08 to 2.49 Å for Ce-N, and 1.80 to 2.29 Å for Hf-N. Notably, the characteristic vibrational frequencies observed in Raman spectroscopy correlate well with the interatomic distances obtained from PDF, both reflecting variations in bond lengths[51,52]. The metal-N stretching vibrations exhibit a blue shift relative to the metal-Cl peaks in Raman spectroscopy, consistent with the shorter metal-N first-neighbor distances observed in the PDF analysis. Also, the local coordination details of polyhedra in $x$$Li_3N$-$MCl_y$, revealed by AIMD simulations, highlight the role of nitrogen in promoting the formation of amorphous phases (Fig. 3d). In crystalline $Li_2MgCl_4$, $LiCeCl_4$, and $Li_2HfCl_6$, rigid polyhedral units ($[MgCl_6]^{4-}$, $[CeCl_8]^{5-}$, $[HfCl_6]^{2-}$) maintain fixed coordination numbers. In contrast, their amorphous counterparts, with nitrogen incorporation, form dynamic bridged networks with fluctuating coordination numbers, ranging from 4–6 in 0.1$Li_3N$-$MgCl_2$, 5–8 in 0.3$Li_3N$-$CeCl_3$, and 4–7 in 0.533$Li_3N$-$HfCl_4$ (Supplementary Fig. 10). In short, $Li_3N$ drives amorphization through two synergistic mechanisms. First, by partially substituting Cl, nitrogen introduces coordination asymmetry and forms distorted polyhedra in $x$$Li_3N$-$MCl_y$, disrupting short-range structural periodicity. Second, $N^{3-}$ acts as a bridge, linking adjacent polyhedra via corner- or edge-sharing configurations to generate random amorphous networks. Collectively, Raman, AIMD, and PDF analysis converge on a single conclusion: $N^{3-}$ acts as the key driver, proliferating a wide range of network formers. Its elevated charge density and pronounced polarizability diversify both coordination geometries and inter-polyhedral linkages, generating a highly cross-linked yet aperiodic framework that substantially elevates AFA.

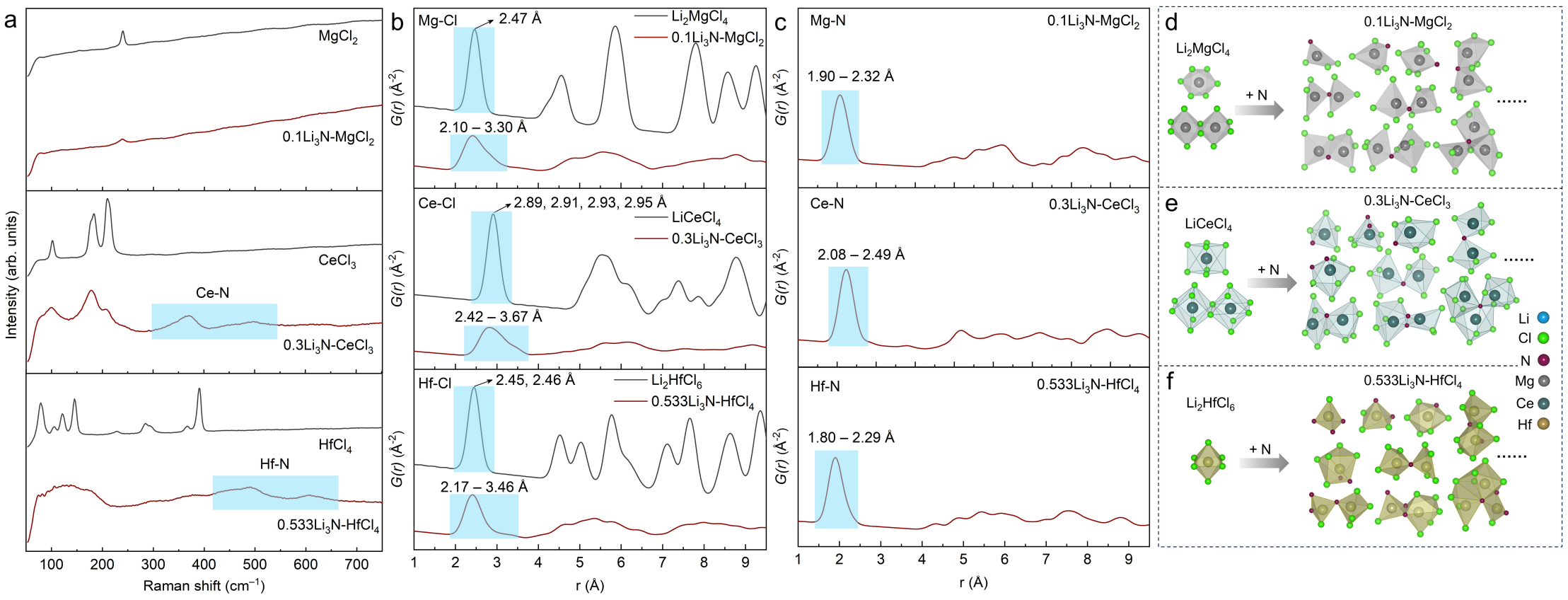

**Figure 3. Local structure analysis of amorphous $x$Li$_3$N-MCl$_y$ SSEs.** (a) Raman spectra of pristine metal chlorides versus amorphous $x$Li$_3$N-MCl$_y$. (b) PDF of amorphous $x$Li$_3$N-MCl$_y$ and crystalline Li$_A$MCl$_B$ ($A$ = 1 or 2; M = Mg, Ce, Hf; $B$ = 4 or 6). (c) PDF of metal-N for amorphous $x$Li$_3$N-MCl$_y$. (d) Coordination polyhedra of crystalline Li$_A$MCl$_B$ and amorphous $x$Li$_3$N-MCl$_y$ based on metal-centred clusters.

**Dynamic Li$^+$ transport in amorphous $x$Li$_3$N-MCl$_y$ SSEs.**

Ionic conductivities of $x$Li$_3$N-MCl$_y$ were evaluated through electrochemical impedance spectroscopy (EIS; Supplementary Figs. 11 and 12). 0.1Li$_3$N-MgCl$_2$, 0.3Li$_3$N-CeCl$_3$, and 0.533Li$_3$N-HfCl$_4$ exhibit the conductivities of $3.97 \times 10^{-4}$ mS cm$^{-1}$, 0.0217 mS cm$^{-1}$, and 2.02 mS cm$^{-1}$, respectively, at 25 °C (Fig. 4a). Their electronic conductivities, measured by direct current (DC) methods, are $2.30 \times 10^{-9}$, $2.30 \times 10^{-9}$, and $1.19 \times 10^{-8}$ S cm$^{-1}$ (Supplementary Fig. 13), respectively, confirm the dominance of ionic conduction in $x$Li$_3$N-MCl$_y$ SSEs. Nyquist and Arrhenius plots for $x$Li$_3$N-MCl$_y$, shown in Supplementary Figs. 14 and 15 illustrate the temperature dependence of ionic conductivity. The activation energies, derived from the Arrhenius equation $\sigma = A \exp(-E_a/k_BT)$, where $\sigma$ denotes ionic conductivity, $A$ the preexponential parameter, $E_a$ the activation energy, $T$ the absolute temperature, and $k_B$ the Boltzmann constant, decrease from 0.572 eV in 0.1Li$_3$N-MgCl$_2$ to 0.346 eV in 0.3Li$_3$N-CeCl$_3$ and 0.307 eV in $0.533Li_3N$-$HfCl_4$ (Fig. 4b). At 1200 K, the mean-squared displacement (MSD) analysis based on AIMD simulations reveals that the $Li^+$ diffusion coefficients are $3.48 \times 10^{-5}$, $9.33 \times 10^{-5}$, and $1.37 \times 10^{-4}$ $cm^2$ $s^{-1}$ for $0.1Li_3N$-$MgCl_2$, $0.3Li_3N$-$CeCl_3$, and $0.533Li_3N$-$HfCl_4$, respectively, showing a gradual increase consistent with experimental ionic conductivity trends (Fig. 4c). For $0.533Li_3N$-$HfCl_4$, temperature-dependent diffusion coefficients (Supplementary Fig. 16) yield an MSD-derived activation energy of 0.335 ± 0.032 eV, closely matching the experimental value of 0.307 eV.

PDF analysis, coupled with structure modeling, elucidates the relationship between local structure and $Li^+$ transport in amorphous $x Li_3N$-$MCl_y$ SSEs (Fig. 4d-g). First, the disordered network in amorphous $x Li_3N$-$MCl_y$ supports a "dynamic monkey bar" mechanism[21,31], where $Li^+$ migrates through transient bonding with $Cl^-$. The incorporation of higher-valence cations (e.g., $Hf^{4+}$ in $0.533Li_3N$-$HfCl_4$) favors the formation of $Cl^-$-rich amorphous networks, thereby enhancing $Li^+$ mobility by providing more dynamic pathways. Second, $Li^+$ concentration is a key factor in determining ionic conductivity. In the $x Li_3N$-$MCl_y$ system, a polymer-like dissociation mechanism allows the amorphous matrix to release free $Li^+$ from $Li_3N$[30]. Higher-valence cations form more M–N bonds, which facilitate more $Li^+$ release, leading to higher ionic conductivity in $0.533Li_3N$–$HfCl_4$. Third, shorter distances between neighboring Li sites typically facilitate ion transport by enhancing pathway connectivity. Partial PDF analysis (Fig. 4d) reveals a gradual contraction of Li-Li spacing from 2.99 Å ($0.1Li_3N$-$MgCl_2$) to 2.47 Å ($0.3Li_3N$-$CeCl_3$) and 2.32 Å ($0.533Li_3N$-$HfCl_4$), which aligns with the observed increase of ionic conductivity. Last but not least, the amorphous $0.533Li_3N$-$HfCl_4$ exhibits a more compact disordered network. PDF analysis (Fig. 4f) shows a shorter Hf–N distance (1.90 Å) compared to Hf–O (2.02 Å)[53,54], due to stronger covalent Hf–N bonding[55]. The higher bridging capability of $N^{3-}$ also leads to shorter Hf–Hf and Cl–Cl distances, indicating enhanced network densification and a continuous $Li^+$ channel that favors $Li^+$ transporting (Supplementary Table 1)[18,19,37,56-58].

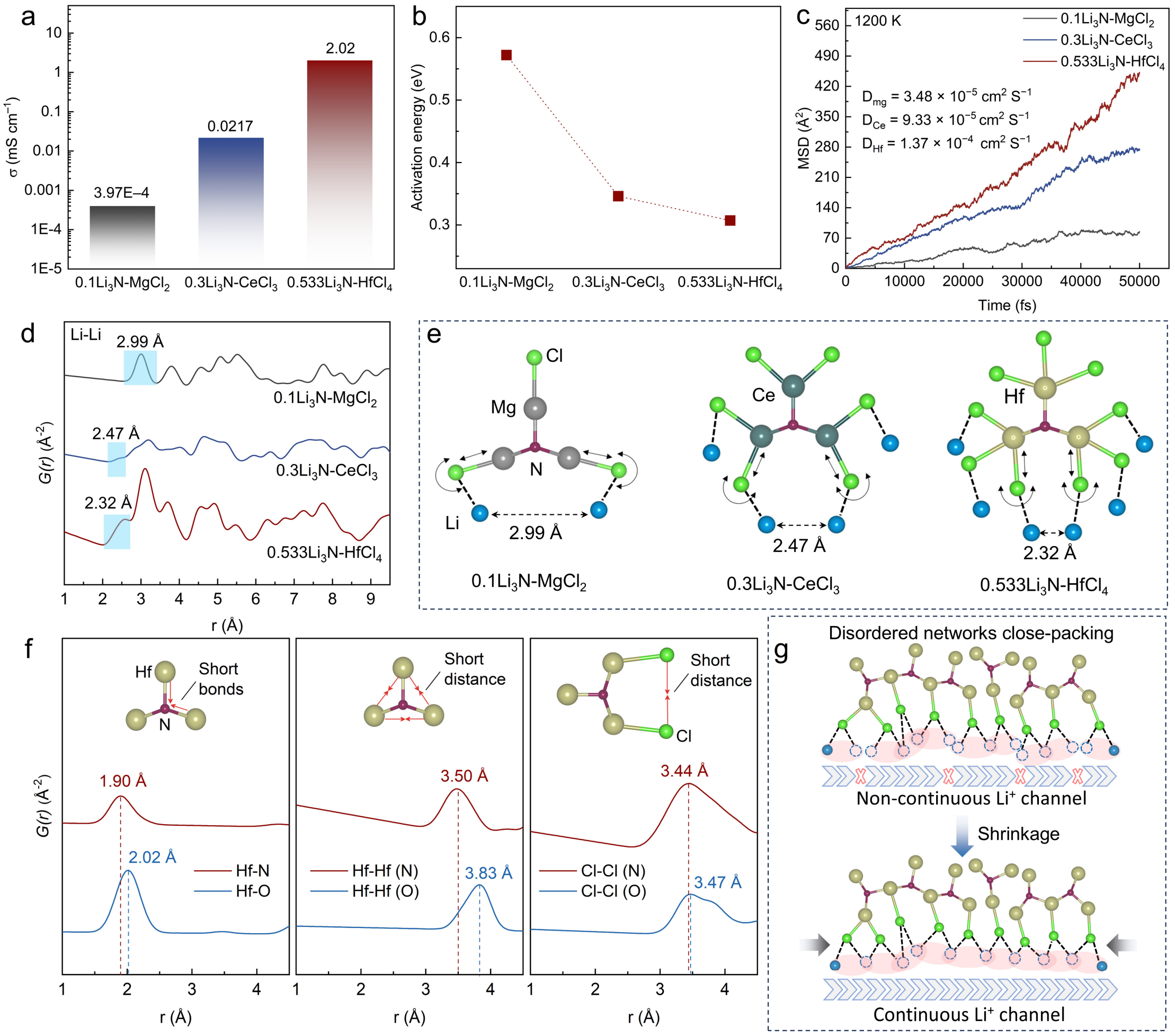

**Figure 4. Ionic transport properties, structural dynamics, and morphological features of amorphous $x$Li3N-MCl$_y$ SSE.** (a) Ionic conductivities of amorphous $x$Li3N-MCl$_y$ at 25 °C. (b) The activation energies of amorphous $x$Li3N-MCl$_y$. (c) MSD and diffusivities of $Li^+$ in $x$Li3N-MCl$_y$ at 1200 K. (d) PDF of Li-Li based on amorphous $x$Li3N-MCl$_y$. (e) Schematic of the N-centered short-range structure in $x$Li3N-MCl$_y$. (f) PDF of Hf-N/O, Hf-Hf, and Cl-Cl based on amorphous N-Cl or O-Cl based SSEs. The

PDF data of O-Cl based SSE from the Sun group's previous work[12]. (g) Schematic of $Li^+$ diffusion mechanism involving N-bridged networks and variable Cl coordination.

**High-Voltage All-Solid-State Batteries with Unprecedented Stability**

The amorphous $0.533Li_3N$-$HfCl_4$ SSE, which shows the highest ionic conductivity of 2.0 mS $cm^{-1}$ at RT, was selected to evaluate the electrochemical performance in high-voltage ASSBs. These ASSBs demonstrate good interfacial stability, rate capability, and cycling durability under diverse practical conditions. ASSBs configurations employing Li-In anodes and high-loading cathodes ($LiCoO_2$ (LCO) and $LiNi_{0.83}Mn_{0.05}Co_{0.12}O_2$ (NCM 83), ~6.24 mg $cm^{-2}$) demonstrated ultralow area resistance (<20 Ω $cm^2$ at 25 °C), as confirmed by Nyquist analysis (Fig. 5a), indicative of superior ionic transport kinetics and negligible charge-transfer barriers, indicating favorable application prospects[59]. This remarkable performance is attributed to the tailored SSE bilayer design: the amorphous $0.533Li_3N$-$HfCl_4$ layer establishes intimate contact with oxide cathodes, promoting efficient charge transfer, as evidenced by energy-dispersive spectroscopy (EDS) results showing uniform coating of $0.533Li_3N$-$HfCl_4$ on the cathode particles surface and effective filling of interparticle voids (Supplementary Fig. 17), while the $Li_{5.5}PS_{4.5}Cl_{1.5}$ (LPSC) layer suppresses deleterious side reactions at the Li-In interface. Rate capability tests underscored outstanding high-current resilience, with the Li-In|LPSC|$0.533Li_3N$-$HfCl_4$|LCO cell delivering a discharge capacity of 135 mAh $g^{-1}$ after switching from 910 mA $g^{-1}$ to 28 mA $g^{-1}$ within a potential range of 2.5–4.2 V vs. $Li^+$/Li (Fig. 5b,c) and the Li-In|LPSC|$0.533Li_3N$-$HfCl_4$|NCM83 cell achieving 180.3 mAh $g^{-1}$ upon transitioning from 1000 mA $g^{-1}$ to 40 mA $g^{-1}$ within a potential range of 2.5–4.3 V vs. $Li^+$/Li (Fig. 5d,e). Long-term cycling tests further highlighted the robustness of electrolyte, with the LCO-based ASSB retaining 83.29% capacity (105 mAh $g^{-1}$) after 200 cycles at 140 mA $g^{-1}$ within a potential range of 2.5–4.2 V vs.

$Li^+/Li$, and preserving 84.06% capacity (91.3 mAh $g^{-1}$) after 300 cycles at 280 mA $g^{-1}$ (Fig. 5f, g and Supplementary Fig. 18 and 19). Strikingly, the NCM83 configuration exhibited unparalleled stability at ultrahigh rates of 1000 mA $g^{-1}$, maintaining discharge capacities of 98.9, 89.0, and 82.7 mAh $g^{-1}$ after 1000, 2000, and 3000 cycles, respectively, corresponding to 90.98%, 81.87%, and 76.05% capacity retention (Fig. 5h and Supplementary Fig. 20). Compared to previously reported SSEs[19,37,56,57] (Supplementary Fig.21 and Supplementary Table 2), the ASSBs based on amorphous $0.533Li_3N\text{-}HfCl_4$ SSE demonstrate distinct advantages in rate capability and long-term cycling stability, underscoring the critical role of the amorphous $0.533Li_3N\text{-}HfCl_4$ SSE in enabling robust interfacial dynamics and high ionic conductivity.

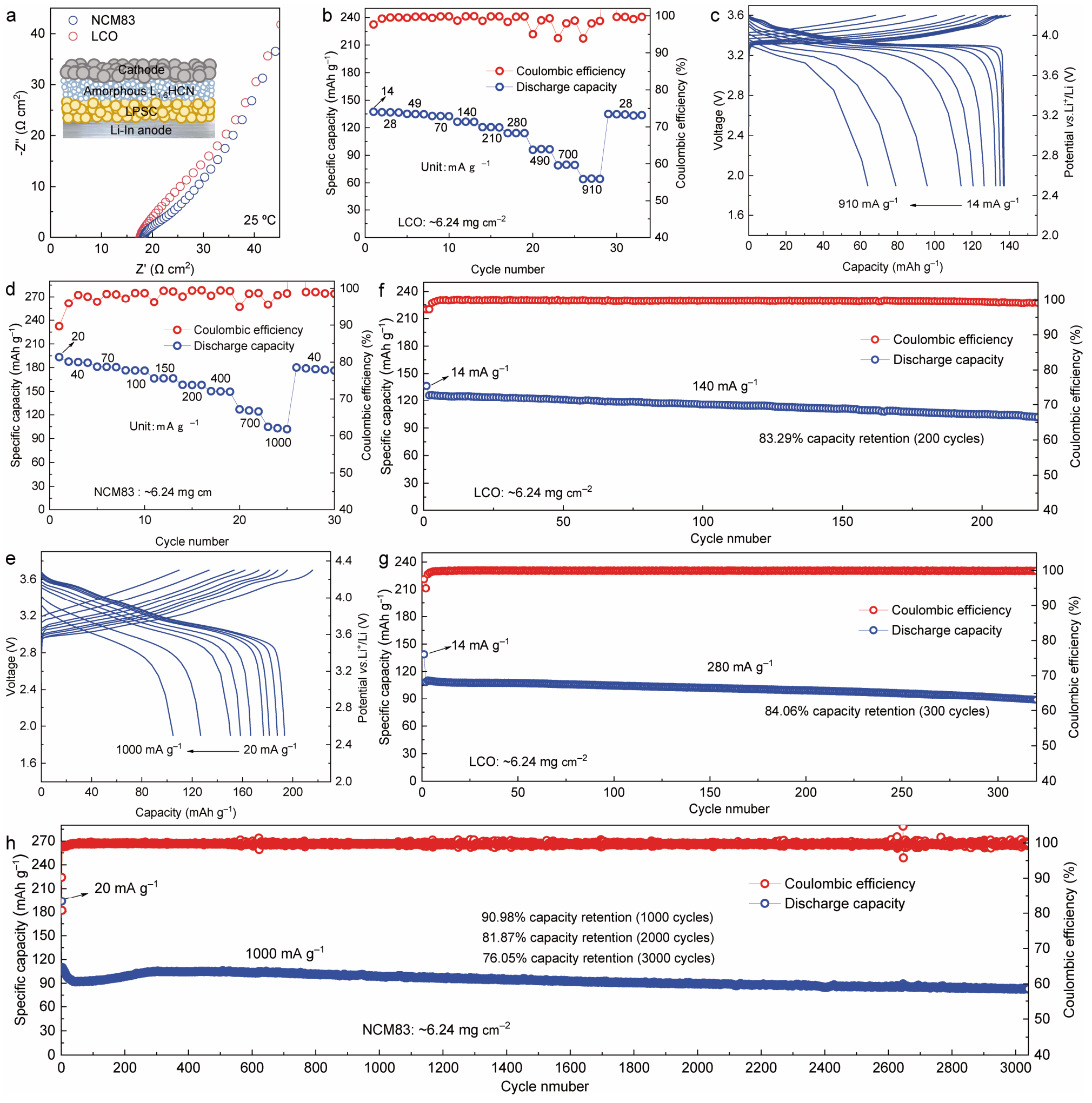

**Figure 5. Electrochemical performance of ASSBs based on amorphous $0.533Li_3N$-**

**$HfCl_4$ SSE.** (a) Nyquist plots of Li-In|LPSC|0.533$Li_3N$-$HfCl_4$|LCO/NCM83 ASSBs under 25 °C (insert: schematic of the ASSBs). (b-e) The cycling performance of ASSBs at 27 °C and various rates. Li-In|LPSC|0.533$Li_3N$-$HfCl_4$|LCO from 14 to 910 mA $g^{-1}$ with a charging/discharging potential range of 2.5-4.2 V vs. $Li^+$/Li (b, c), and Li-In|LPSC|0.533$Li_3N$-$HfCl_4$|NCM83 from 20 to 1000 mA $g^{-1}$ with a charging/discharging potential range of 2.5-4.3 V vs. $Li^+$/Li (d, e). (f, g) The cycling performance of LCO|0.533$Li_3N$-$HfCl_4$-LPSC|Li-In with a charging/discharging potential range of 2.5-4.3 V vs. $Li^+$/Li under 27 °C at 140 (f) and 280 (g) mA $g^{-1}$. (h) The cycling performance Li-In|LPSC|0.533$Li_3N$-$HfCl_4$|NCM83 with a charging/discharging potential range of 2.5-4.3 V vs. $Li^+$/Li under 27 °C at 1000 mA $g^{-1}$.

## Methods

### Material synthesis

Amorphous $x$$Li_3N$-$MCl_y$ ($0.3 \leq 3x \leq 1.9$; M denotes a metal element; $2 \leq y \leq 4$) SSEs were synthesized via mechanochemical processing under an argon atmosphere. $Li_3N$ (>99.4%, Alfa Aesar) and metal chlorides ($HfCl_4$, $ZrCl_4$, $CeCl_3$, $YCl_3$, $MgCl_2$, $MnCl_2$; >99.9%, Macklin or Aladdin) were mixed in stoichiometric molar ratios corresponding to the target composition. The resulting mixture was loaded into zirconia jars (45 mL) with zirconia milling balls (5–10 mm) at a 45:1 ball-to-powder mass ratio and processed using a planetary ball mill (Fritsch Pulverisette 7). The material was first low-speed ball milling at 100 rpm for 3 hours, followed by high-energy ball milling at 500 rpm for 30 hours. The milling protocol employed an intermittent bidirectional rotation pattern, with 15 minutes of counterclockwise rotation, a 5-minute pause, and 15 minutes of clockwise rotation, repeated throughout the process. All handling and

milling operations were conducted under continuous argon protection to prevent atmospheric contamination.

**Material characterizations and analysis**

Structural characterization of the synthesized $x$$Li_3N$-$MCl_y$ SSE was performed using multiple analytical techniques. Amorphous phase identification was executed via X-ray diffraction analysis (Empyrean, Malvern Panalytical) employing monochromatic Cu Kα radiation (λ = 1.5406 Å), with specimens encapsulated within Kapton polyimide membranes under inert conditions to prevent air exposure. The tests were conducted at a scan rate of 4.5° $min^{-1}$ from 10 to 90°. Morphological evaluation and elemental mapping were conducted through field-emission scanning electron microscopy (SEM, JEOL-JSM7610) coupled with a cryogenically cooled silicon drift detector for energy-dispersive X-ray spectroscopy (EDS). The XCT measurements were conducted at Nanovoxel 3432E (Sanying, China). The visualization and segmentation of XCT images were performed using Avizo software (Thermo Fisher Scientific, USA). Chemical bonding was analyzed using a confocal Raman microscope (Horiba LabRAM Odyssey) equipped with a 532 nm laser excitation, employing argon-filled quartz capillaries (1.0 mm internal diameter) as sample containment vessels. The cryogenic transmission electron microscopy (cryo-TEM) characterizations were carried out using an aberration-corrected Titan Krios TEM operated at 300 kV. The $^7$Li solid-state NMR (SSNMR) experiments were performed on a Bruker Avance NEO 400 spectrometer equipped with a 9.4 T wide-bore magnet using a 3.2 mm HXY double-resonance MAS probe. The $^7$Li MAS NMR spectra were acquired using a one-pulse sequence with a π/2 pulse length of 3.0 μs and a recycle delay of 2.0 s at a spinning rate of 10 kHz. $^{7}$Li chemical shifts with respect to a 1 M LiCl solution at 0.0 ppm.

**Construction of amorphous structures and pair distribution function simulation**

The crystalline structure of $Li_2MgCl_4$ was obtained from the Inorganic Crystal Structure Database (ICSD)[43], while those of $LiCeCl_4$ and $Li_2HfCl_6$ were retrieved from the Open Quantum Materials Database (OQMD)[44]. Ab initio molecular dynamics (AIMD)[60] simulations were then performed in the NPT[61] ensemble to thermally treat these crystalline structures. The systems were gradually heated to 1500 K, inducing a transition from the ordered crystalline phase to a fully molten state. Following melting, rapid quenching was applied to generate the corresponding amorphous structures. The melt-quench protocol was carried out under a Langevin thermostat, with a total simulation time of 10 ps and a time step of 1 fs. The resulting amorphous configurations are presented in Supplementary Fig. S9.

In this study, AIMD simulations were performed on the constructed amorphous structures within the NVT[62] ensemble, with the system temperature regulated by a Nosé-Hoover thermostat. The simulations were conducted with a time step of 1 fs for a total duration of 50 ps. All AIMD calculations were carried out using the Vienna Ab initio Simulation Package (VASP)[60], employing the projector augmented-wave (PAW)[63] method together with the Perdew-Burke-Ernzerhof (PBE)[64] generalized gradient approximation (GGA)[65] for the exchange-correlation functional. A plane-wave energy cutoff of 340 eV was used, and the Brillouin zone was sampled at the Γ point. The total number of atoms in the $0.1Li_3N$-$MgCl_2$, $0.3Li_3N$-$CeCl_3$, and $0.533Li_3N$-$HfCl_4$ systems was 148, 208, and 228, respectively. In addition, the pair correlation functions *G(r)* were calculated using PDFgui software, based on CIF files generated from AIMD simulations[66].

**AIMD data analysis: ionic diffusion and activation energy**

For $0.1Li_3N{\cdot}MgCl_2$, $0.3Li_3N{\cdot}CeCl_3$, and $0.533Li_3N{\cdot}HfCl_4$, AIMD simulations were conducted at 800, 1000, 1200, 1300, and 1500 K to examine ionic transport properties. The mean square displacement (MSD) of ions was calculated as:

$$\mathrm{MSD}(t) = \langle |\boldsymbol{r}(t) - \boldsymbol{r}(0)|^2 \rangle$$

where $r(t)$ and $r(0)$ denote the positions of a particle at time $t$ and at the initial time, respectively, and $\langle \cdot \rangle$ represents an ensemble average over all particles and time origins. After equilibration, the systems entered a steady-state diffusion regime, during which the MSD increased linearly with time. The diffusion coefficient D was extracted from the slope according to:

$$D = \lim_{t \to \infty} \frac{1}{2dt} \langle |\boldsymbol{r}(t) - \boldsymbol{r}(0)|^2 \rangle$$

where $d$=3 for three-dimensional systems. The temperature dependence of $D$ was then analyzed using the Arrhenius relation:

$$D(T) = D_0 \exp\left(-\frac{E_a}{k_B T}\right)$$

where $D(T)$ is the diffusion coefficient at temperature T, $D_0$ is the pre-exponential factor reflecting the intrinsic structural and dynamic characteristics of the system, $E_a$ is the activation energy for diffusion, $k_B$ is Boltzmann's constant ($8.617\times10^{-5}$ eV/K), and T is the absolute temperature. This relationship highlights the exponential enhancement of $D$ with increasing temperature and the crucial role of $E_a$ in governing ionic mobility.

By plotting ln$D$ versus $\frac{1}{T}$ (Arrhenius plots) and performing linear fits, the slope $-\frac{E_a}{k_B}$ was obtained, allowing quantitative evaluation of the activation energies and thus the ionic transport behavior across the studied systems.

**Electrochemical measurements and fabrication of batteries**

The ionic conductivity of synthesized $x$$Li_3N$-$MCl_y$ SSEs was quantitatively assessed through EIS measurements conducted under potentiostatic mode. The $x$$Li_3N$-$MCl_y$ SSEs powder (100 mg) was prepared by cold-pressing within a polyether-ether-ketone (PEEK) model cell (10 mm inner diameter) under 2 tons with two stainless steel rods as blocking electrodes, producing compacted pellets with controlled thicknesses (0.4–0.7 mm). EIS test using a Solartron 1470E impedance analyzer, with measurements spanning a frequency range of 0.1 Hz–1 MHz (10 mV voltage amplitude) at different temperatures. Each impedance spectrum comprised 71 discrete frequency points, from which ionic conductivity (σ) was calculated via the following equation:

$$\sigma = \frac{L}{R\pi r^2}$$

where $L$, $R$, and $r$ denote pellet thickness, ohmic resistance from Nyquist plot extrapolation, and electrode radius, respectively. The cell assembly process for DC measurements was similar to EIS tests. To determine the electronic conductivity, the current responses of the cell were measured at 0.2 V for 1800 s.

A multi-layered architecture was engineered for solid-state battery construction through sequential powder compaction processes under an inert atmosphere (<0.1 ppm $H_2O$/$O_2$). The composite cathode formulation consisted of $LiCoO_2$ (LCO, from Guangdong Canrd New Energy Technology Co., Ltd.) or $LiNi_{0.83}Co_{0.12}Mn_{0.05}O_2$ (NCM83, from GEM Co., Ltd.), amorphous $0.533Li_3N\text{-}HfCl_4$ electrolyte, and vapor-grown carbon nanofibers (VGCF) in a 70:27:3 mass ratio, homogenized through an agate mortar. Cell assembly using a PEEK mold (10 mm inner diameter). First, 65 mg of $0.533Li_3N\text{-}HfCl_4$ SSE layer (65 mg, 1 ton) was evenly spread in the mold, followed by 35 mg of $Li_{5.5}PS_{4.5}Cl_{1.5}$ (LPSC) layered on the $0.533Li_3N\text{-}HfCl_4$ surface. The bilayer electrolyte was consolidated under 2 tons of pressure. Subsequently, 7 mg of composite cathode was uniformly spread on the $0.533Li_3N\text{-}HfCl_4$ side of the bilayer pellet and pressed at 3 tons. The anode was formed by attaching high-purity indium foil (φ10 mm) and lithium foil (φ8 mm, weight ratio Li/In = 1:40) to the LPSC layer. Finally, stainless steel current collectors were then inserted at both ends of the cell, and the assembly was completed by pressing at 0.5 tons to optimize interfacial contact. All electrochemical performance tests were performed using the Land and Neware battery testing system within temperature-controlled chambers to maintain isothermal operating conditions.

## Resource availability

### Lead contact

Further information and requests for resources should be directed to and will be fulfilled by the lead contact, Prof. J. Zhu (zhujl@sustech.edu.cn).

### Materials availability

All unique reagents generated in this study are available upon reasonable request.

### Data and code availability

Information required to reanalyze the data reported in this paper is available from the lead contactupon request.

## Acknowledgments

This study was supported by the Guangdong Grants (2021ZT09C064) and Guangdong Basic and Applied Basic Research Foundation (2024B1515120042 ), the National Natural Science Foundation of China (12426301, 12275119, 52227802, 12405343), Shenzhen Science and Technology Program (KQTD20200820113047086.), Shenzhen Key Laboratory of Solid State Batteries (ZDSYS20180208184346531), Guangdong-Hong Kong-Macao Joint Laboratory for Photonic-Thermal-Electrical Energy Materials and Devices(2019B121205001), and Guangdong Provincial Key Laboratory of Energy Materials for Electric Power (2018B030322001), we also acknowledge the Major Science and Technology Infrastructure Project of Material Genome Big-science Facilities Platform supported by the Municipal Development and Reform Commission of Shenzhen.

## Author contributions

Conceptualization: Conceptualization: B. Hong, L. Gao, S. Han, J. Zhu. Methodology: B. Hong, L. Gao, M. Liu, S. Han, J. Zhu. Visualization: B. Hong, L. Di, H. Ni, P. Nan, Y. Li, Z. Lei. Funding acquisition: S. Han, J. Zhu. Project administration: S. Han, J. Zhu. Supervision: S. Han, J. Zhu. Writing – original draft: B. Hong, S. Han. Writing – review & editing: L. Gao, S. Han, J. Zhu. All authors contributed to the discussion about the manuscript.

## Declaration of interests

The authors declare no competing financial interests.

## Supplementary information

Supplementary Information is available for this paper at https://doi.org/******

# *Supplementary Information for*

# Nitrogen-Triggered Amorphization Enables High-Performance Solid-State Electrolytes

Bolong Hong[1,2], Lei Gao[3]*, Bingkai Zhang[4], Pengfei Nan[5], Ruishan Zhang[4], Yuhang Li[6], Zhihao Lei[6], Ming Liu[6], Jing Wu[7], Longbang Di[3], Haijin Ni[2], Songbai Han[2]*, Jinlong Zhu[2,8]*

[1]*College of Semiconductors (National Graduate College for Engineers), Department of Physics, Southern University of Science and Technology, Shenzhen 518055, China.*

[2]*Shenzhen Key Laboratory of Solid State Batteries, Guangdong Provincial Key Laboratory of Energy Materials for Electric Power, Guangdong-Hong Kong-Macao Joint Laboratory for Photonic Thermal-Electrical Energy Materials and Devices, Southern University of Science and Technology, Shenzhen 518055, China.*

[3]*School of Advanced Materials, Peking University, Shenzhen Graduate School, Shenzhen 518055, China.*

[4]*School of Chemical Engineering and Light Industry, Guangdong University of Technology, Guangzhou, 510006, China.*

[5]*Information Materials and Intelligent Sensing Laboratory of Anhui Province, Leibniz International Joint Research Center of Materials Sciences of Anhui Province, Institutes of Physical Science and Information Technology, Anhui University, Hefei 230601, China.*

[6]*Shenzhen Geim Graphene Center, Tsinghua University, Shenzhen International Graduate School, Shenzhen 518055, China.*

[7]*Cryo-EM Center, Southern University of Science and Technology, Shenzhen, Guangdong, 518055, China.*

[8]*Department of Physics, Southern University of Science and Technology, Shenzhen 518055, China.*

*Corresponding author

Correspondence to: Lei Gao, Songbai Han, Jinlong Zhu

gaolei2018@pku.edu.cn (L. Gao); hansb@sustech.edu.cn (S. Han); zhujl@sustech.edu.cn (J. Zhu)

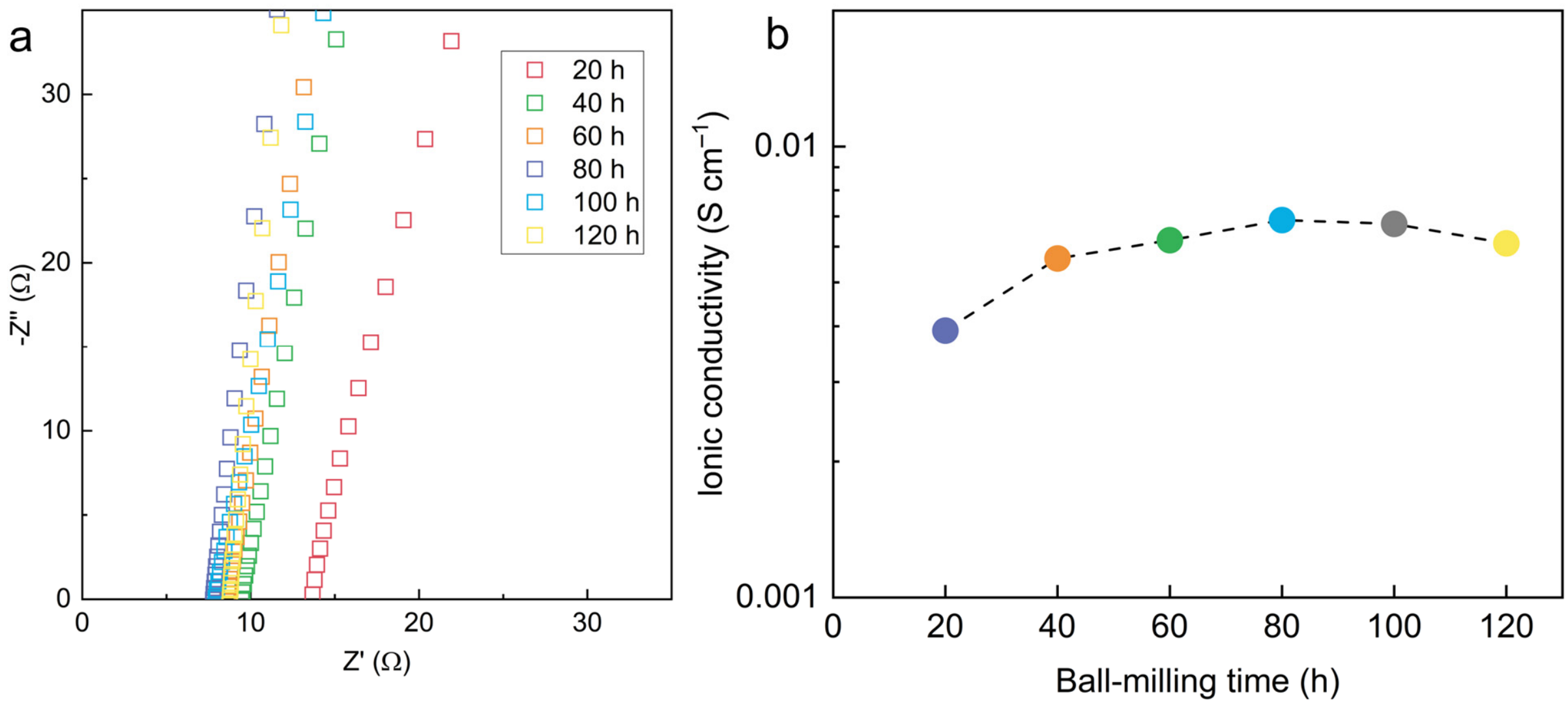

**Supplementary Fig. 1 | Li$^+$ conductivity of amorphous 0.5$Li_2O$-$TaCl_5$ SSE.** (a) Nyquist plots of 0.5$Li_2O$-$TaCl_5$ SSEs under different ball milling conditions at 25 °C. (b) Ionic conductivities of the as-prepared 0.5$Li_2O$-$TaCl_5$ SEs under different ball milling conditions at 25 °C.

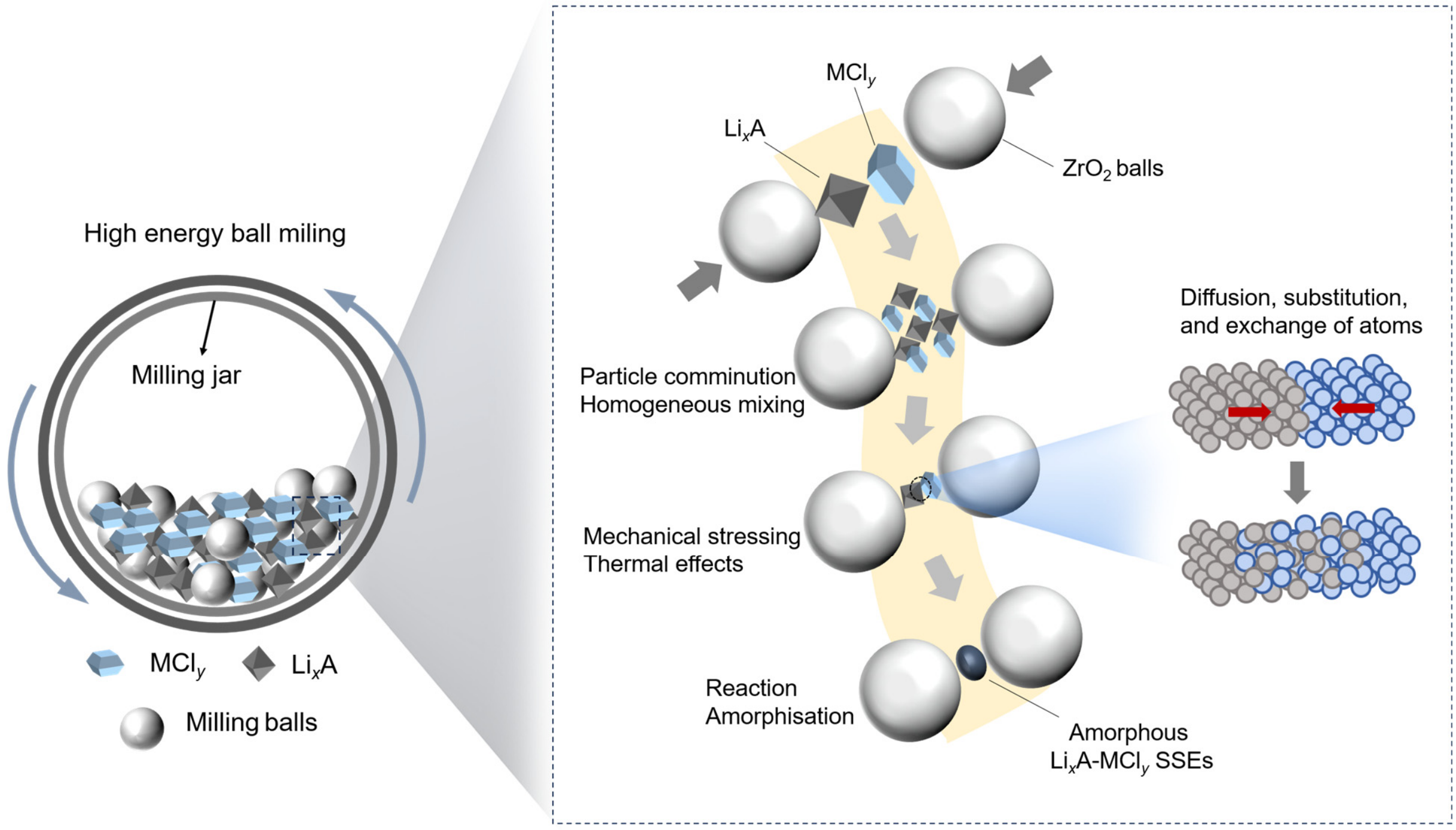

**Supplementary Fig. 2 | Schematic of the synthesis of amorphous $x$Li$_3$A-MCl$_y$ ($x$ = 1, 2, 3; A = Cl, O, N; M denotes a metal element; $y$ = 2, 3, 4, 5) SSEs via high energy ball milling.** The schematic illustrates the mechanochemical process, including particle comminution, atomic exchange between lithium-containing precursors and metal chlorides, and the formation of an amorphous SSE structure[1,2].

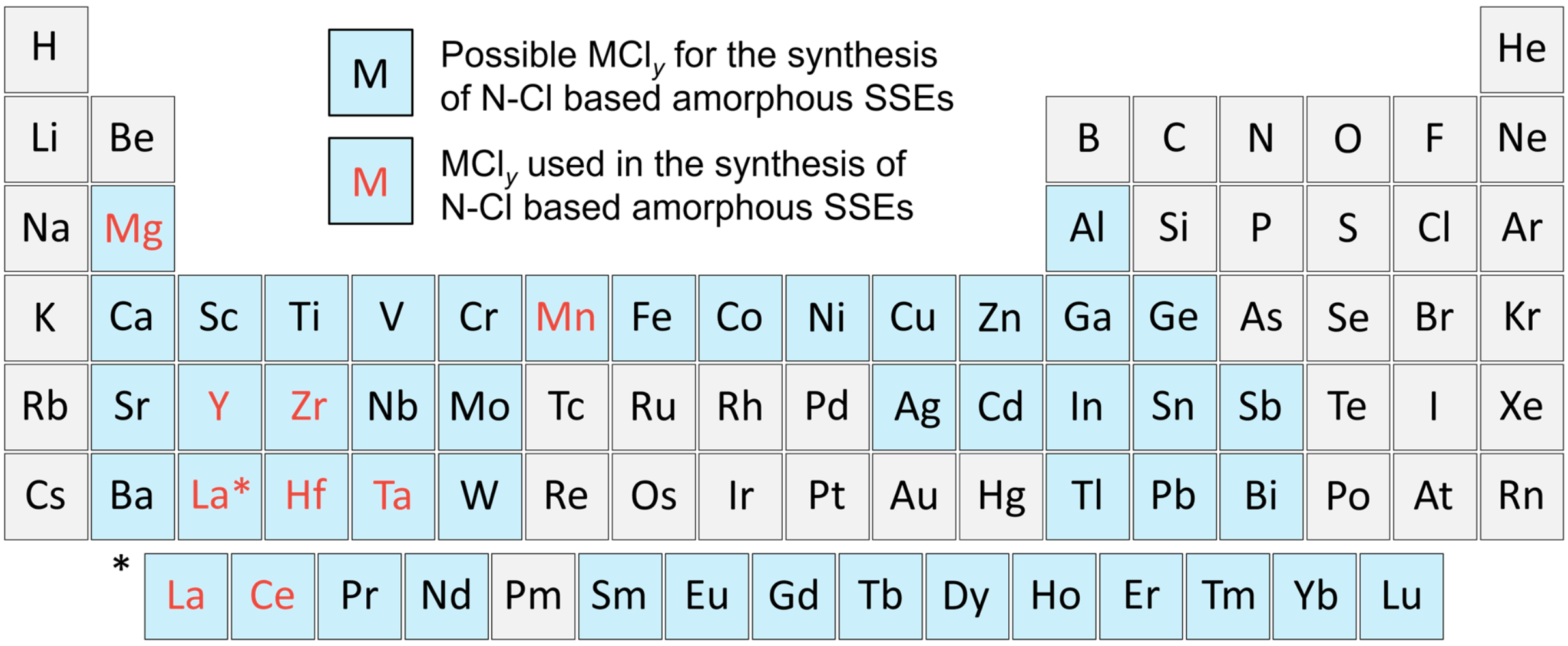

**Supplementary Fig. 3 | A periodic table highlighting the generality of the N-driven synthesis strategy for amorphous SSEs.** Amorphized metal chlorides (e.g., $TaCl_5$, $HfCl_4$) are marked in red, and theoretically compatible candidates (e.g., $AlCl_3$, $FeCl_3$) are in blue. Notably, very rare and radioactive metal elements are excluded.

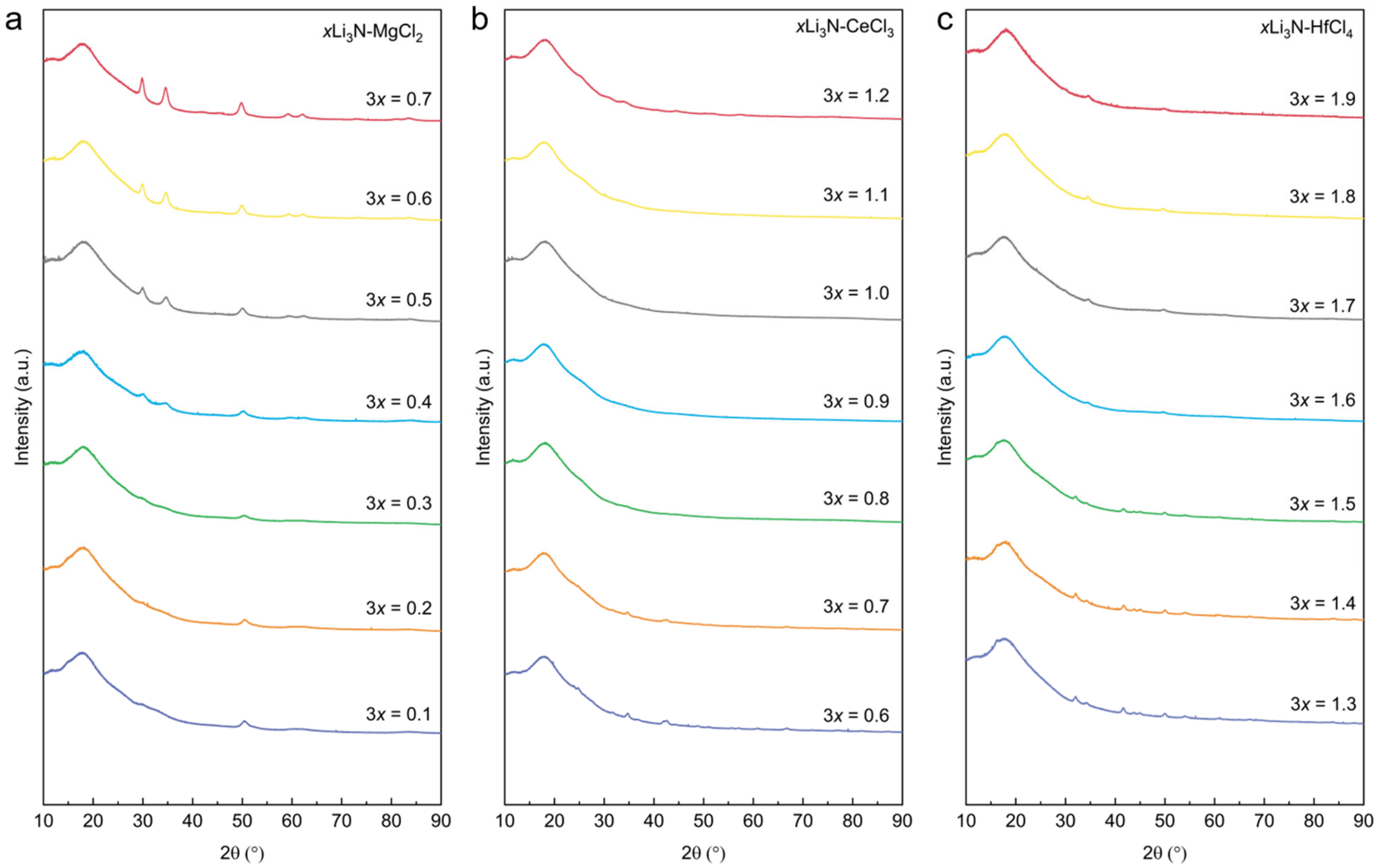

**Supplementary Fig. 4 | X-ray diffraction (XRD) patterns of the amorphous $x$Li$_3$N-MCl$_y$ (0.3 ≤ 3$x$ ≤ 1.9, M = Hf, Ce, Mg, 2 ≤ $y$ ≤ 4). (a)** XRD patterns of $x$Li$_3$N-MgCl$_2$ (0.1 ≤ 3$x$ ≤ 0.7), showing the highest degree of amorphization at 3$x$ = 0.3. **(b)** XRD patterns of $x$Li$_3$N-CeCl$_3$ (0.6 ≤ 3$x$ ≤ 1.2), indicating complete amorphization within 0.8 ≤ 3$x$ ≤ 1.0. **(c)** XRD patterns of $x$Li$_3$N-HfCl$_4$ (1.3 ≤ 3$x$ ≤ 1.9), exhibiting the highest degree of amorphization at 3$x$ = 1.6.

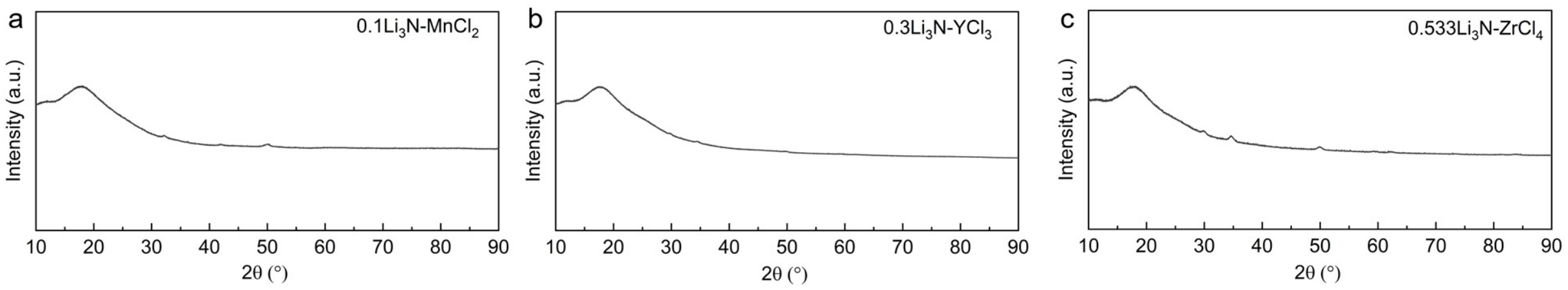

**Supplementary Fig. 5 | (a-c) X-ray diffraction (XRD) patterns of the amorphous *x*$Li_3N$-$MCl_y$ (3*x* = 0.3, 0.9, 1.6, M = Mn, Y, Zr, 2 ≤ *y* ≤ 4). (a)** XRD patterns of the amorphous 0.1$Li_3N$-$MnCl_2$. **(b)** XRD patterns of the amorphous 0.3$Li_3N$-$YCl_3$. **(c)** XRD patterns of the amorphous 0.533$Li_3N$-$ZrCl_4$.

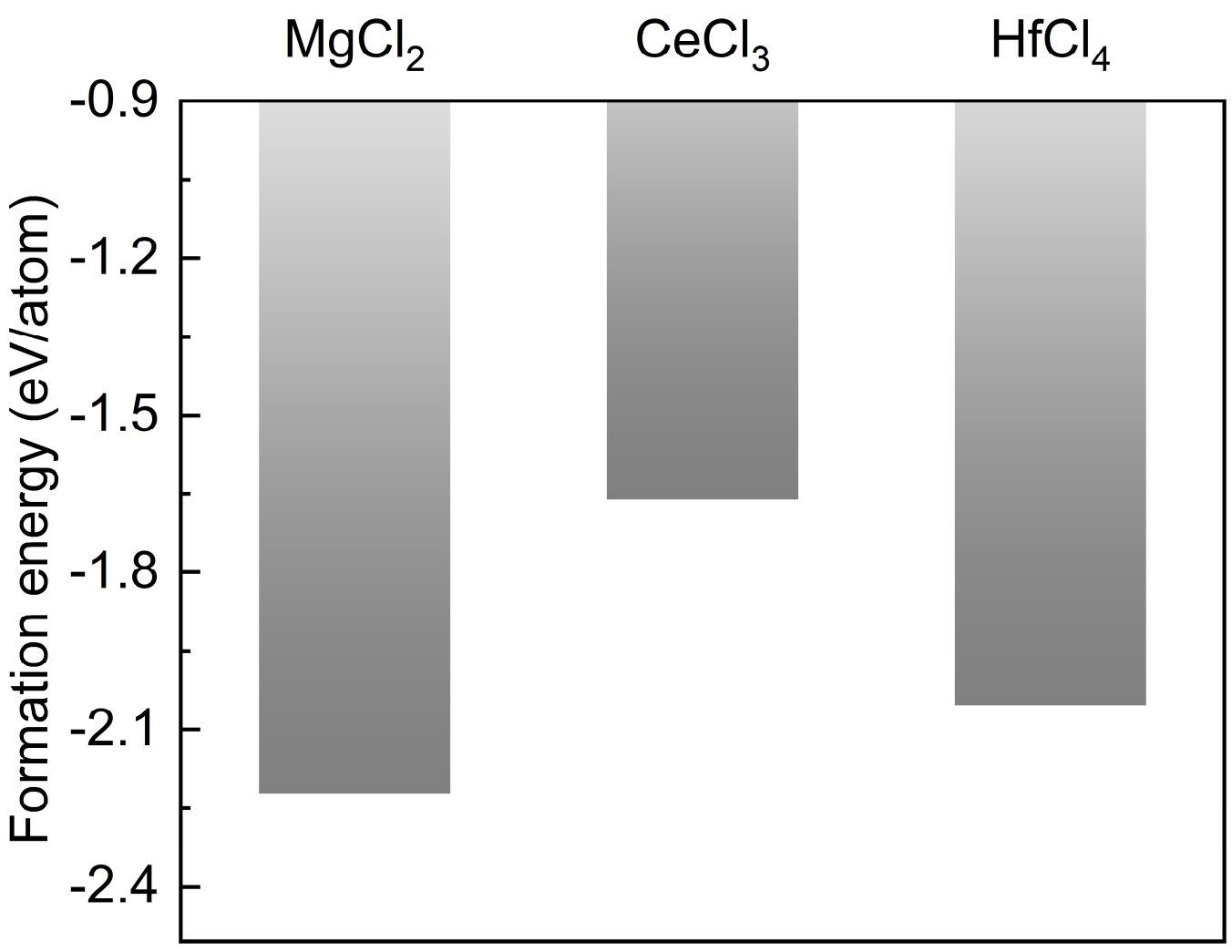

**Supplementary Fig. 6 | Comparison of formation energy of raw materials.** All data were obtained from the Materials Project database[3].

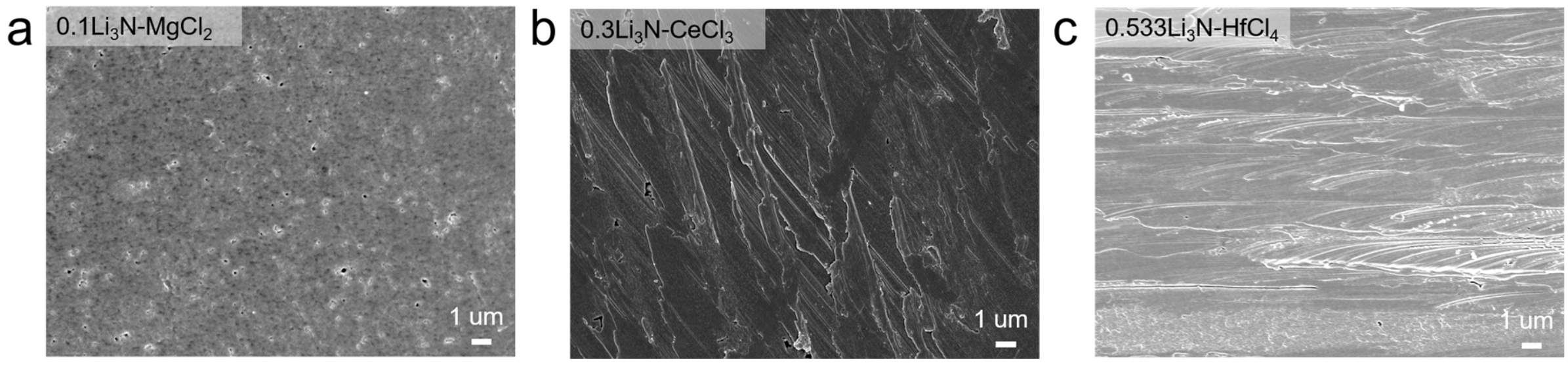

**Supplementary Fig. 7 | (a−c) SEM images of 0.1$Li_3N$-$MgCl_2$ (a), 0.3$Li_3N$-$CeCl_3$ (b), 0.533$Li_3N$-$HfCl_4$ (c) SSEs.** For SEM measurements, SSE pellets were fabricated by cold pressing 60 mg of SSE powder at a pressure of 500 MPa for 30 seconds using a stainless-steel mold with an inner diameter of 6 mm. All processes were carried out in an Ar-filled glove box ($H_2O$, $O_2$ < 0.1 ppm).

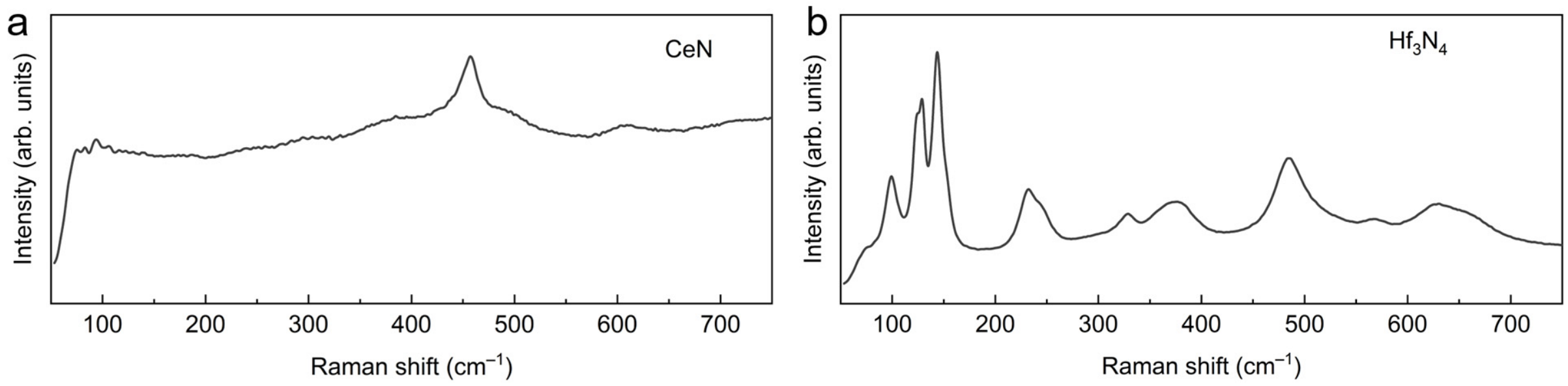

**Supplementary Fig. 8 | (a, b) Raman spectra of CeN (a) and $Hf_3N_4$ (b).**

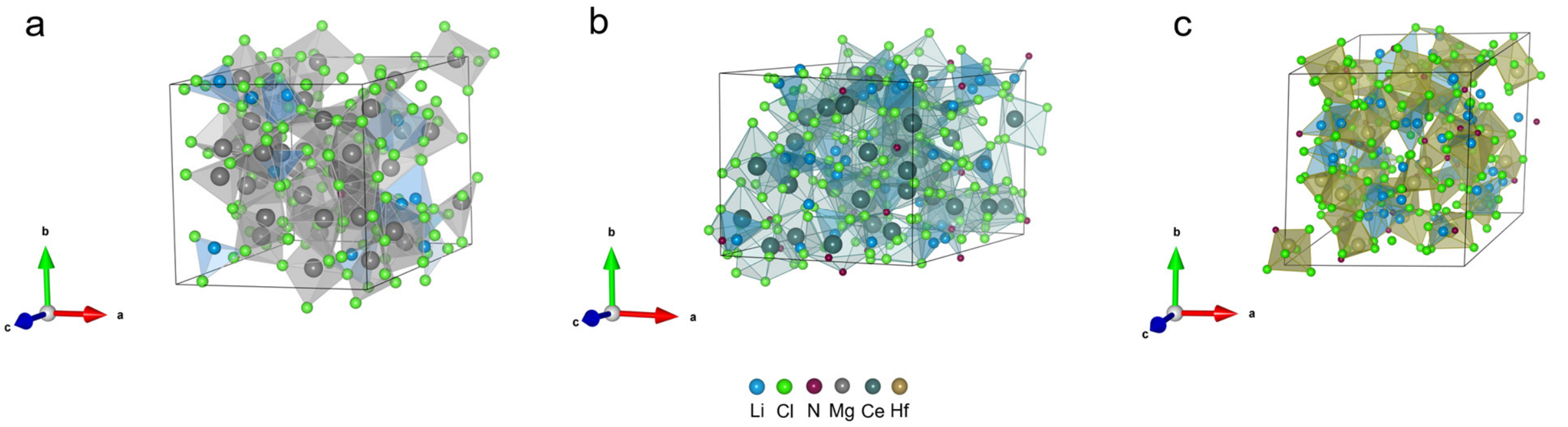

**Supplementary Fig. 9 | (a-c) Computed structure of amorphous $0.1Li_3N$-$MgCl_2$ (a), $0.3Li_3N$-$CeCl_3$ (b), and $0.533Li_3N$-$HfCl_4$ (c) generated from melt-and-quench AIMD simulations.**

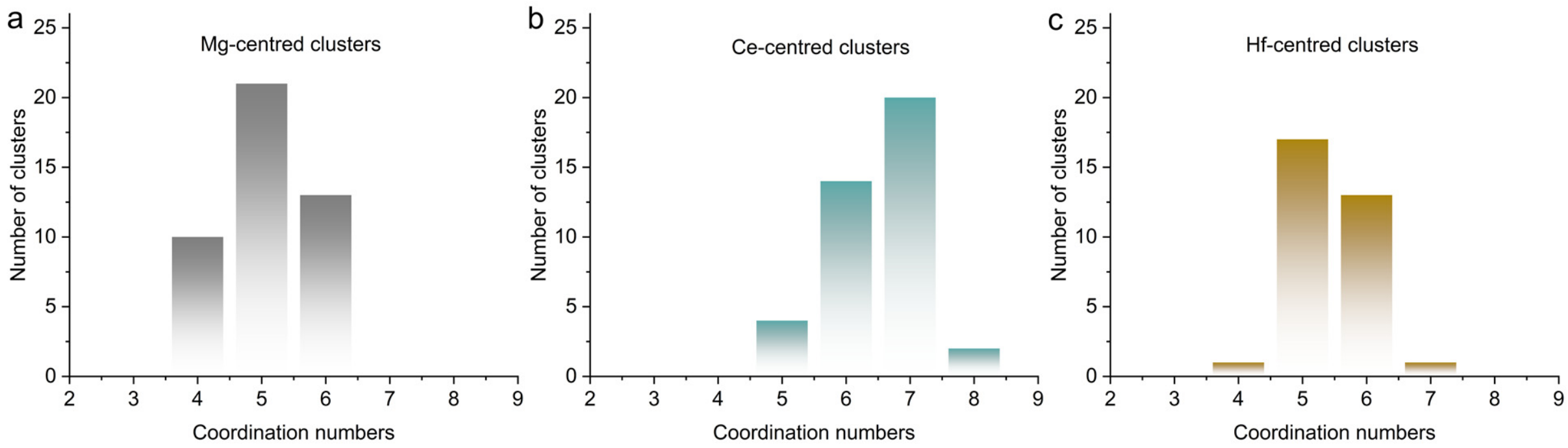

**Supplementary Fig. 10 | (a-c) Relative populations of the Mg-centred (a), Ce-centred (b), and Hf-centred (c) clusters in the computed atomic configuration.**

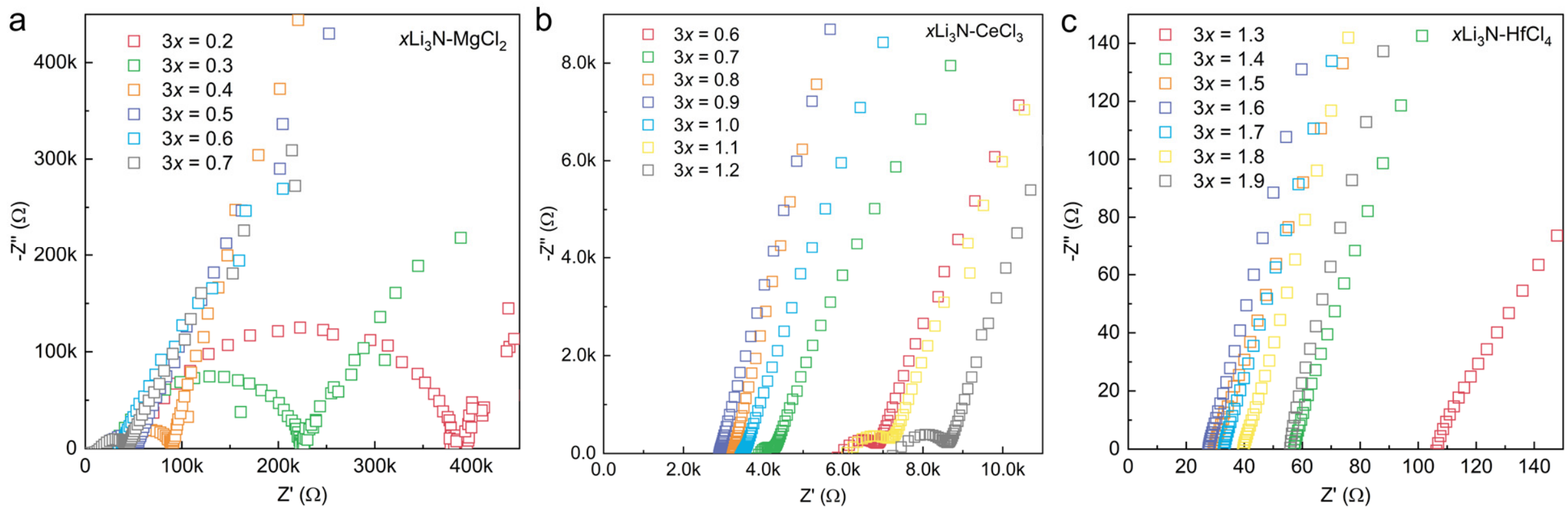

**Supplementary Fig. 11 | (a-c) Nyquist plots reflect the impedance response of amorphous *x*$Li_3N$-$MgCl_2$ (a), *x*$Li_3N$-$CeCl_3$ (b), *x*$Li_3N$-$HfCl_4$ (c) SSEs in various 3*x* values.**

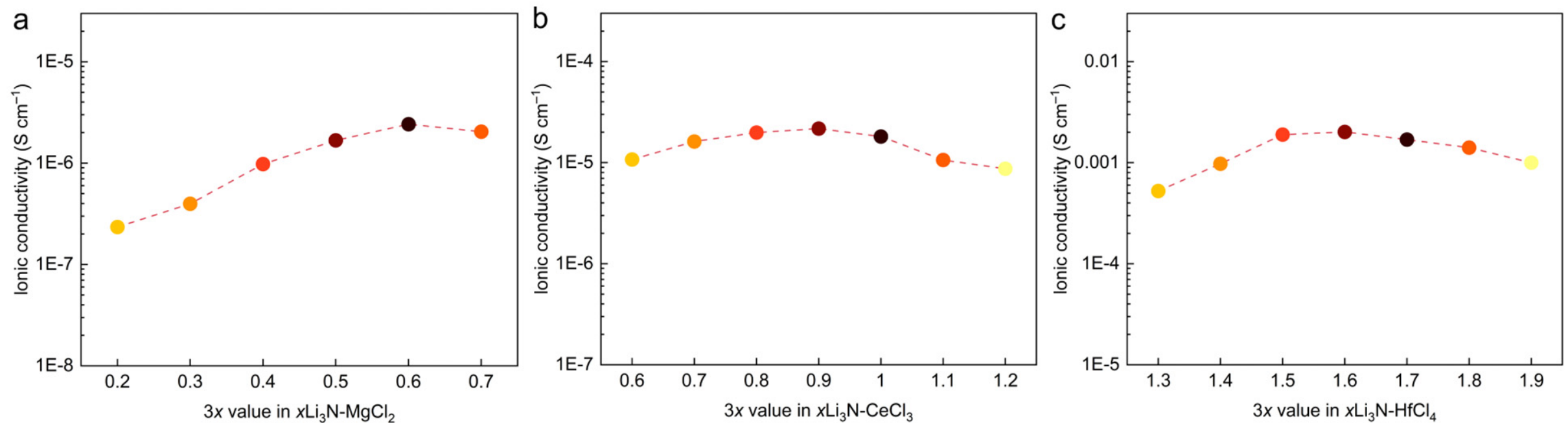

**Supplementary Fig. 12 | (a-c) Ionic conductivities of amorphous 0.1$Li_3N$-$MgCl_2$ (a), 0.3$Li_3N$-$CeCl_3$ (b), and 0.533$Li_3N$-$HfCl_4$ (c) at 25 °C, which were subjected to ball-milling for 30 hours.**

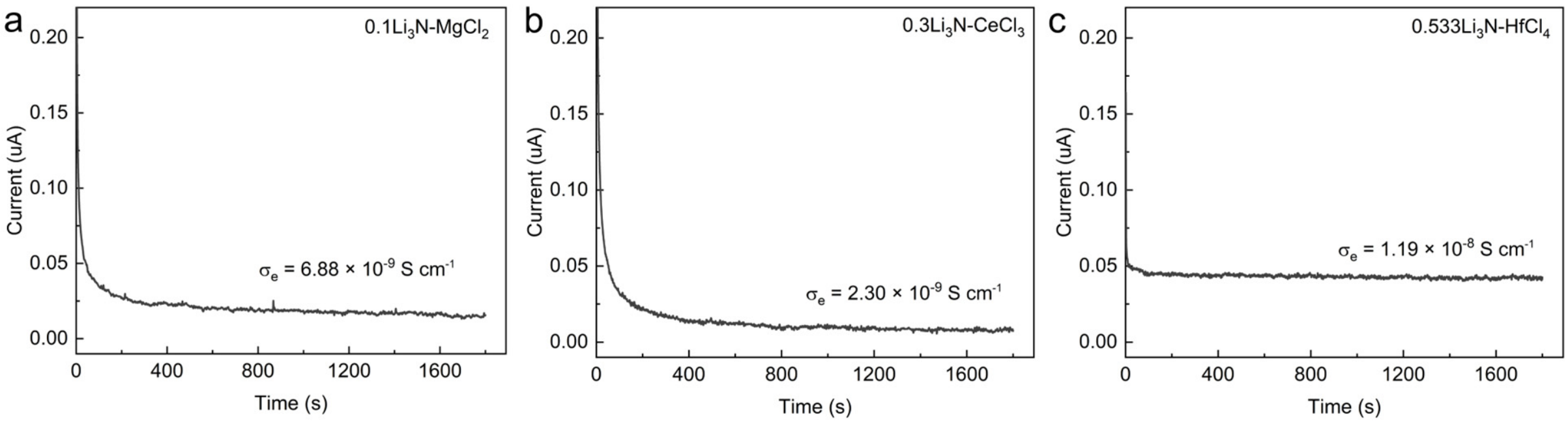

**Supplementary Fig. 13 | (a-c) DC polarization curves of amorphous $0.1Li_3N-MgCl_2$ (a), $0.3Li_3N-CeCl_3$ (b), and $0.533Li_3N-HfCl_4$ (c) SSE using a stainless | SSE | stainless battery at 0.2 V.**

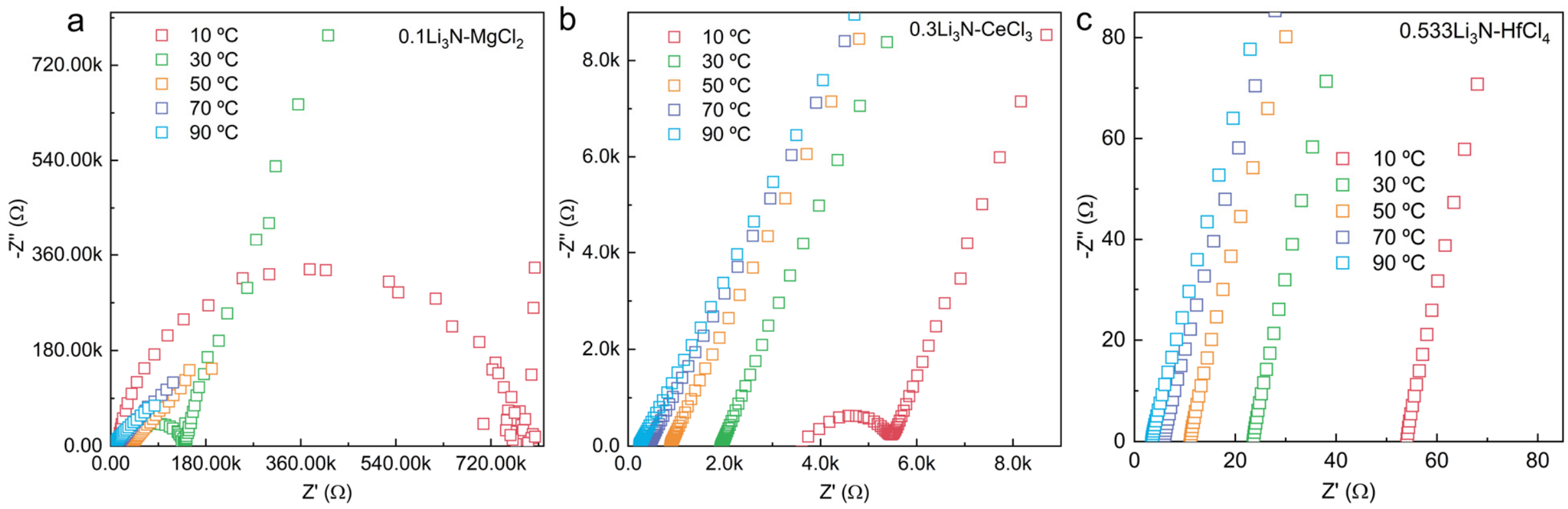

**Supplementary Fig. 14 | (a-c) Nyquist plots reflect the impedance response of amorphous $0.1Li_3N\text{-}MgCl_2$ (a), $0.3Li_3N\text{-}CeCl_3$ (b), $0.533Li_3N\text{-}HfCl_4$ (c) SSEs at the temperature range from 10 to 90 °C.**

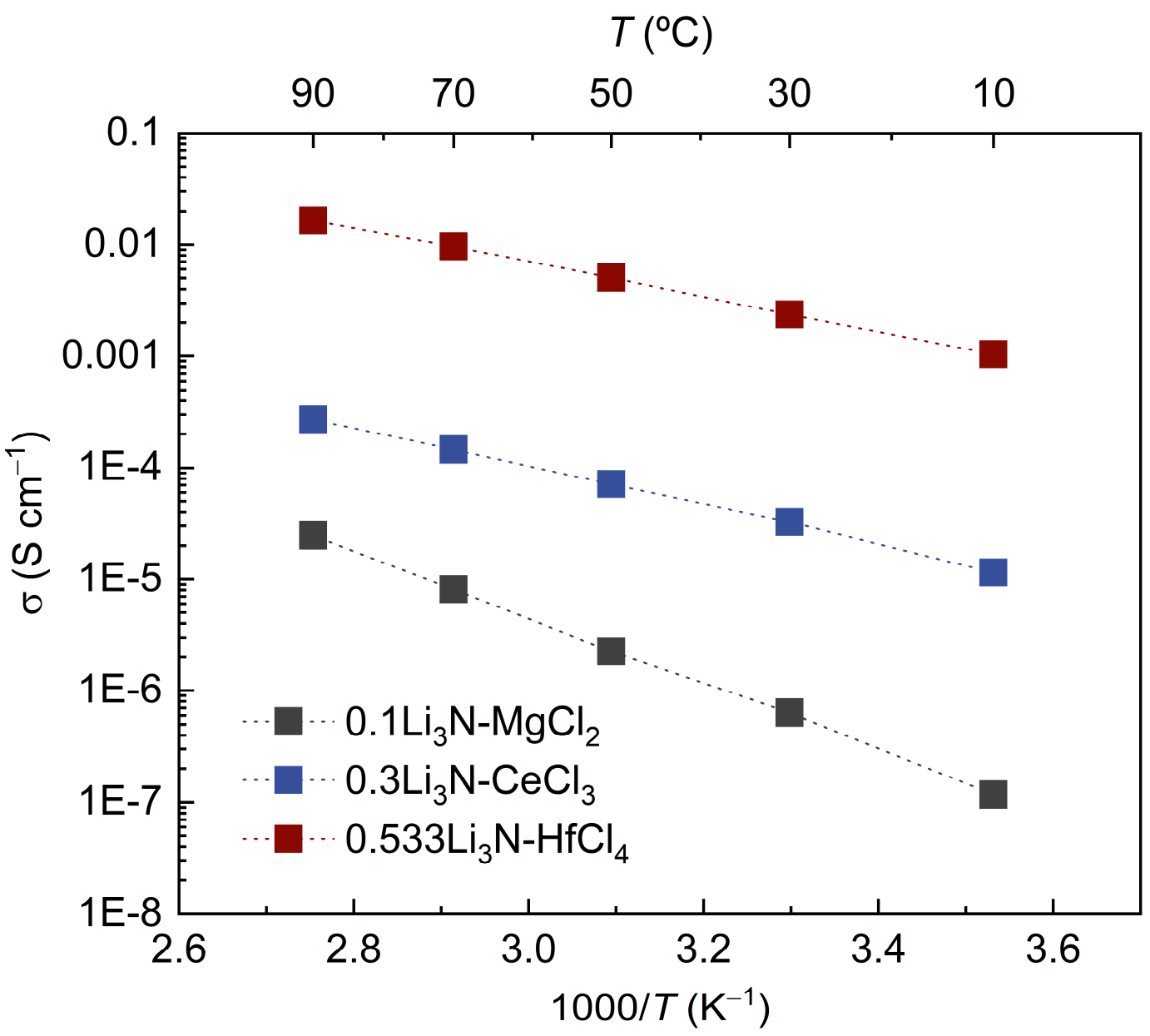

**Supplementary Fig. 15 | Arrhenius ionic conductivity plots of amorphous $x$Li$_3$N-MCl$_y$ in the temperature range from 10 to 90 °C.**

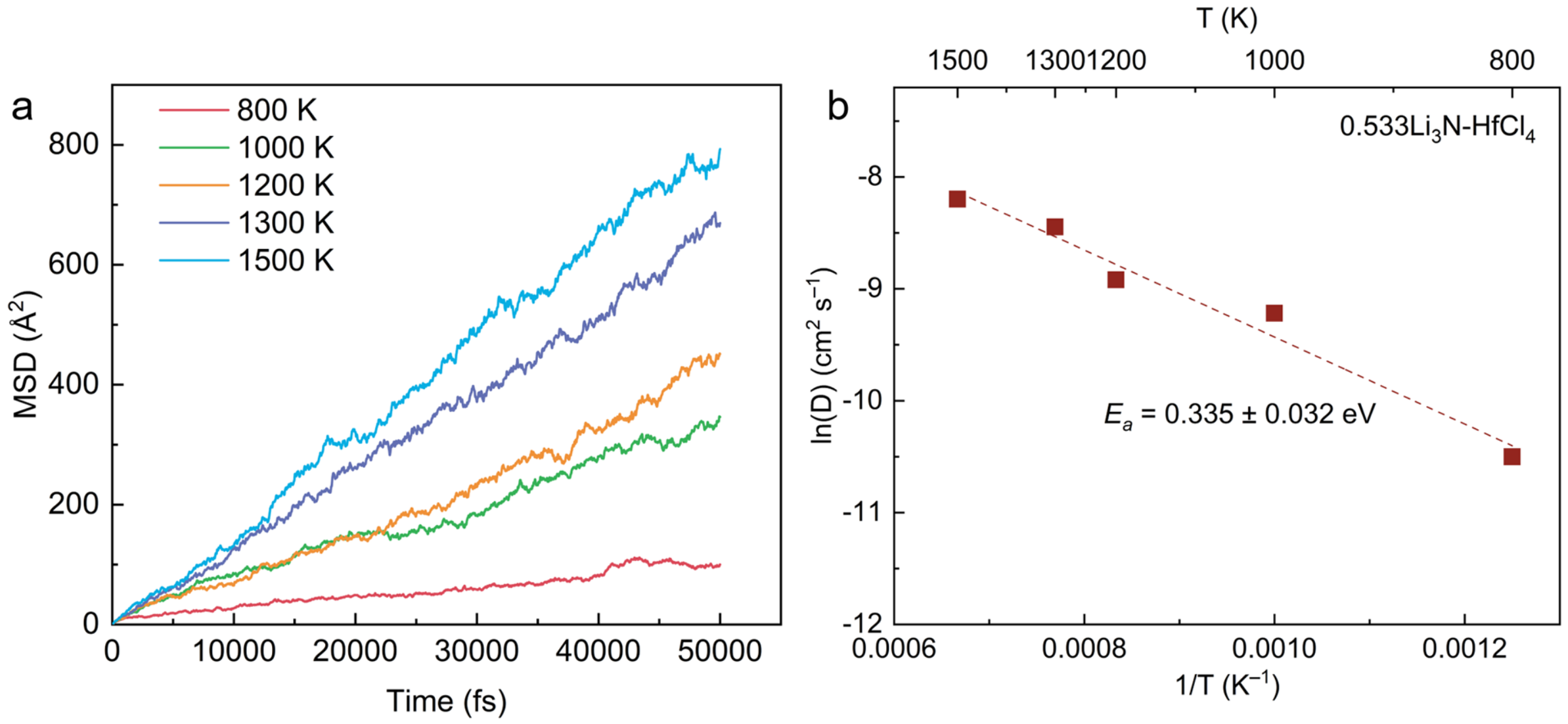

**Supplementary Fig. 16 | Simulated** $Li^+$ **diffusivity, activation energy in 0.533$Li_3N$-$HfCl_4$ from AIMD. (a)** MSD of $Li^+$ in 0.533$Li_3N$-$HfCl_4$ at different temperatures. **(b)** Temperature dependence of $Li^+$ diffusivities in 0.533$Li_3N$-$HfCl_4$ and calculated Arrhenius plots and activation energy.

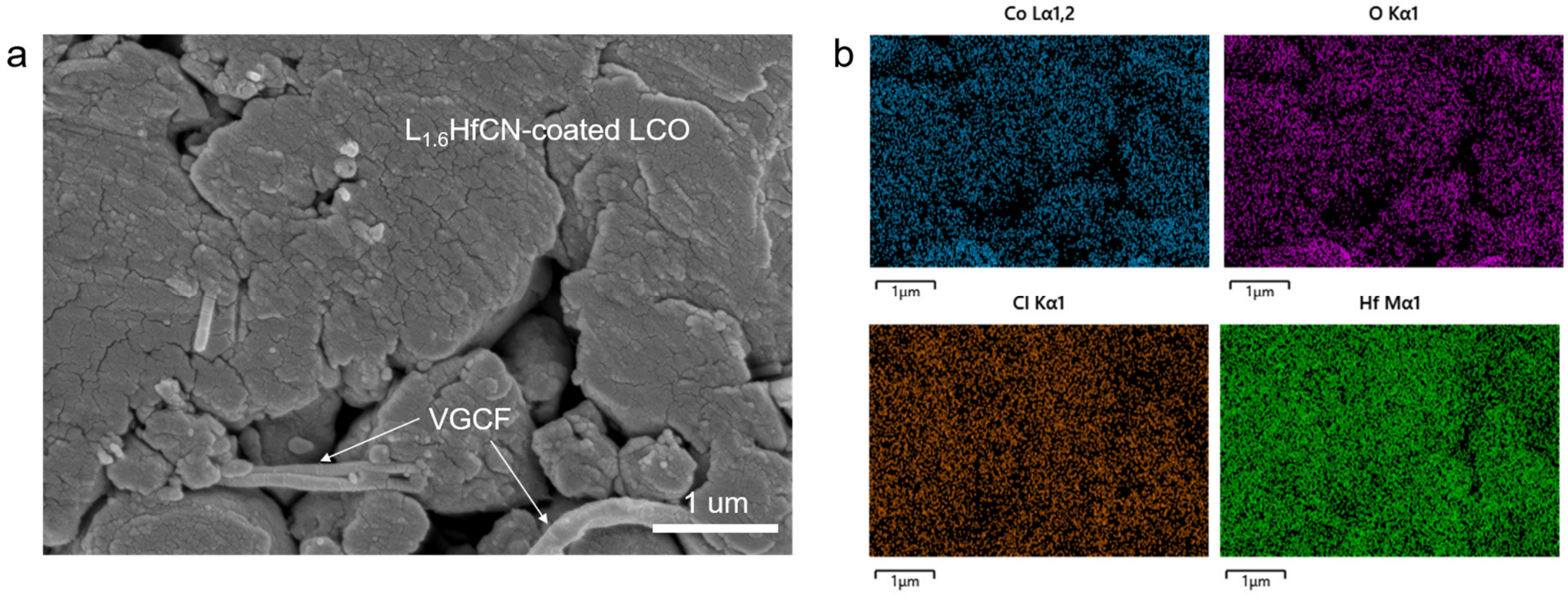

**Supplementary Fig. 17 | Mixed ionic/electronic network within composite cathode. (a)** SEM image for the $0.533Li_3N\text{-}HfCl_4$ coated LCO obtained from simple grinding and cold-pressing. **(b)** EDS mapping of Co, O, Cl, and Hf.

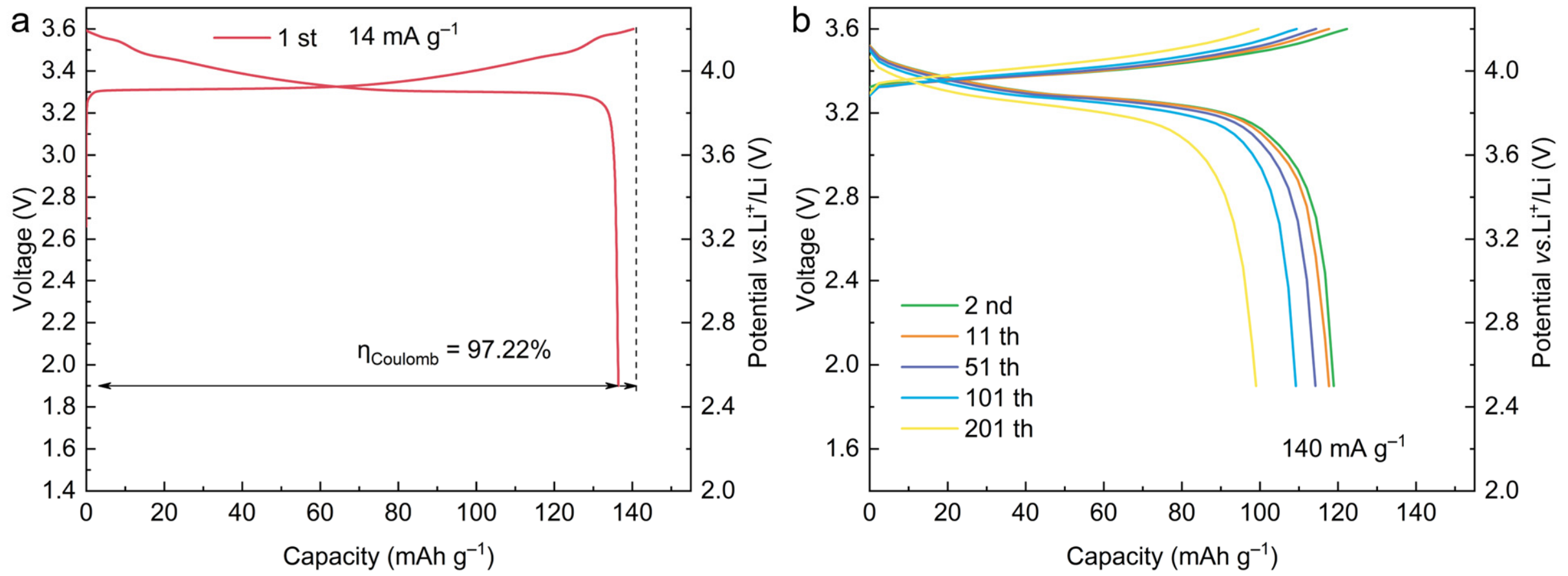

**Supplementary Fig. 18 | Charge/discharge voltage profiles of Li-In|LPSC|0.533$Li_3N$-$HfCl_4$|LCO battery within a potential range of 2.5–4.2 V vs. $Li^+$/Li. (a)** Charge/discharge voltage profiles at 14 mA $g^{-1}$. **(b)** Charge/discharge voltage profiles at different cycles under 140 mA $g^{-1}$.

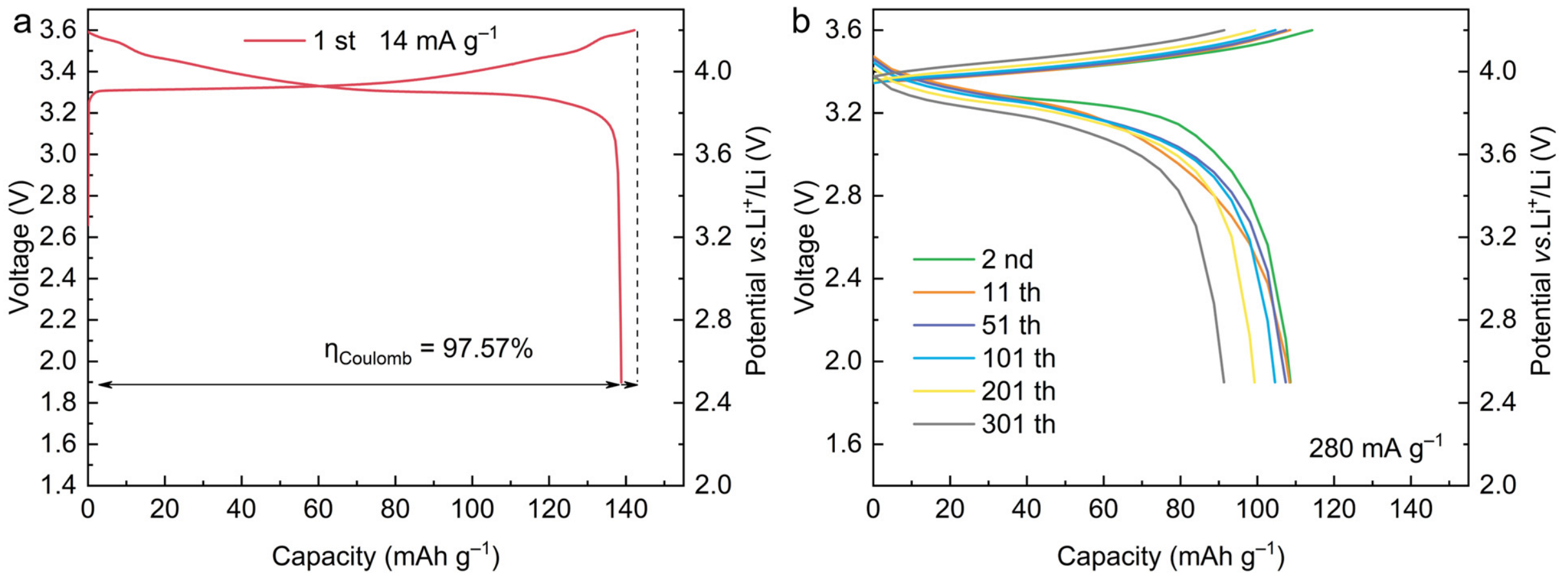

**Supplementary Fig. 19 | Charge/discharge voltage profiles of Li-In|LPSC|0.533$Li_3N$-$HfCl_4$|LCO battery within a potential range of 2.5–4.2 V vs. $Li^+$/Li. (a)** Charge/discharge voltage profiles at 14 mA $g^{-1}$. **(b)** Charge/discharge voltage profiles at different cycles under 280 mA $g^{-1}$.

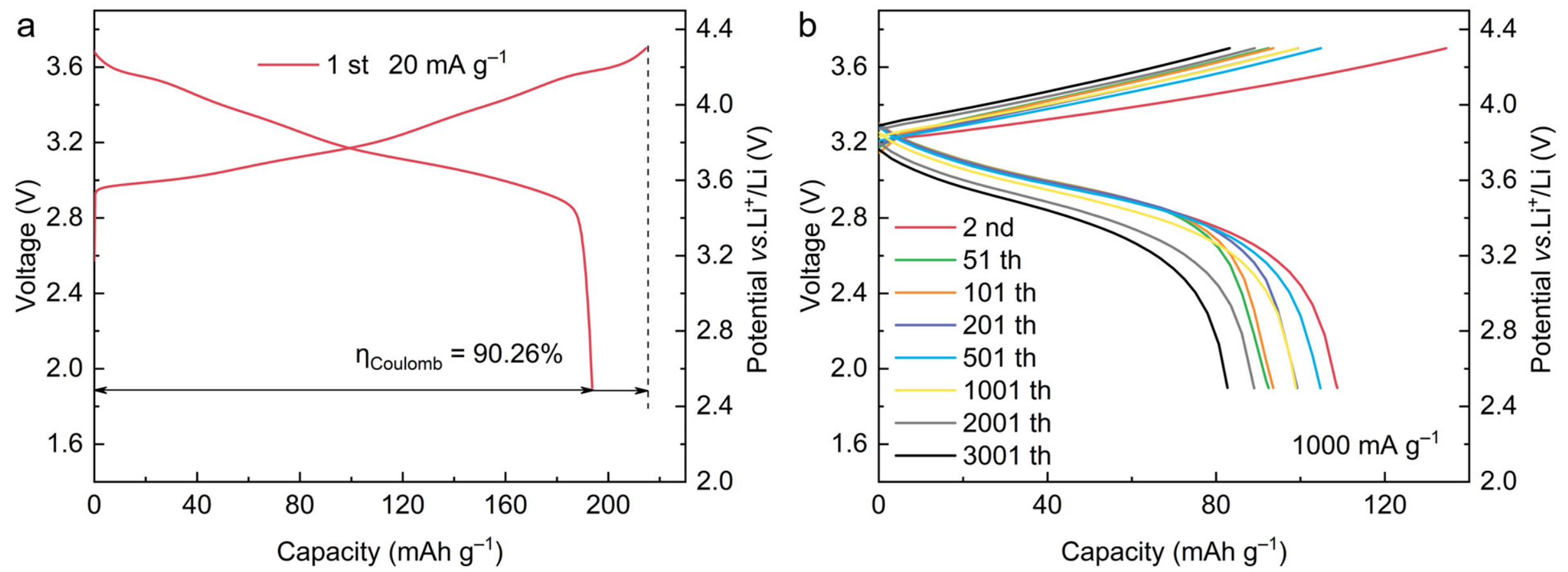

**Supplementary Fig. 20 | Charge/discharge voltage profiles of Li-In|LPSC|0.533$Li_3N$-$HfCl_4$|NCM83 battery within a potential range of 2.5–4.3 V vs. $Li^+$/Li. (a)** Charge/discharge voltage profiles at 20 mA $g^{-1}$. **(b)** Charge/discharge voltage profiles at different cycles under 1000 mA $g^{-1}$.

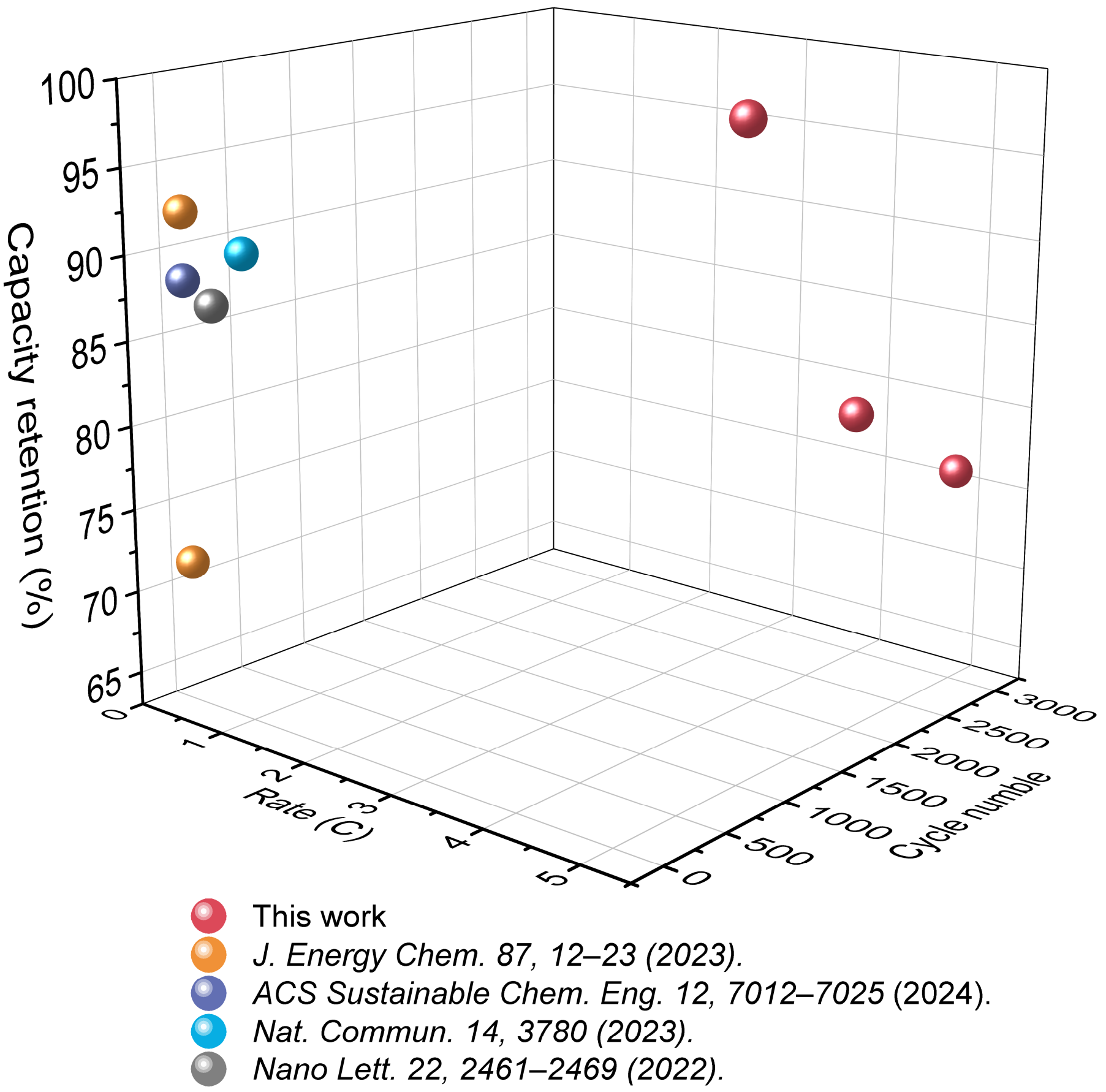

**Supplementary Fig. 21 | Comparison of cycling performance with previously reported results of Hf-based SSEs.**

## Supplementary Table 1
## Ionic conductivity and activation energy comparison for Hf-based SSEs.

| References | Electrolyte | Amorphous/crystalline | Ionic conductivity (mS cm$^{-1}$) | Activation energy (eV) |
|---|---|---|---|---|
| **This work** | **$L_{1.6}HfCN$** | Amorphous | **2.02 (25 °C)** | **0.307** |
| *J. Energy Chem.* ***87**, 12–23 (2023).* | $Li_2HfCl_6$<br>$Li_{2.3}Hf_{0.7}In_{0.3}Cl_6$ | crystalline | 0.398 (30 °C)<br>1.05 (30 °C) | ~0.395<br>0.337 |
| *ACS Sustainable Chem. Eng.* ***12**, 7012–7025 (2024).* | $Li_{2.25}Hf_{0.75}Fe_{0.25}Cl_6$ | crystalline | 0.91 (30 °C) | 0.34 |
| *Nat. Commun.* ***14***, 3780 (2023). | $1.5Li_2O\text{-}HfCl_4$ | Amorphous | 1.97 (25 °C) | 0.328 |
| *Nano Lett.* ***22**, 2461–2469 (2022).* | $Li_2HfCl_6$ | crystalline | ~0.5 (30 °C) | 0.353 |
| *ACS Energy Lett.* ***9**, 1043–1052 (2024).* | $Li_2HfCl_6$<br>$Li_{2.375}Sc_{0.375}Hf_{0.625}Cl_6$ | crystalline | ~0.002 (25 °C)<br>1.1 (25 °C) | ~0.465<br>0.33 |
| *J. Am. Chem. Soc.* ***146**, 2977–2985 (2024).* | $Li_3HfCl_7$<br>$Li_3HfO_{1.5}Cl_4$ | Crystalline<br>Amorphous | 0.3 (25 °C)<br>1.2 (25 °C) | N/A |

## Supplementary Table 2
## Cycling performance comparison for ASSBs.

| References | Electrolyte | Cathode | Anode | Potential range vs.Li/Li$^+$ (V) | Operating temperature (ºC) | Rate (C) | Capacity retention |
|---|---|---|---|---|---|---|---|
| **This work** | **$L_{1.6}HfCN$** | **NCM83** | **Li-In alloy** | **2.5~4.3** | **27** | **5** | **90.98% (1000 cycles)**<br>**81.87% (2000 cycles)**<br>**76.05% (3000 cycles)** |
| *J. Energy Chem.* ***87****, 12–23 (2023).* | $Li_2HfCl_6$<br>$Li_{2.3}Hf_{0.7}In_{0.3}Cl_6$ | $LiCoO_2$ | Li-In alloy | 2.5~4.2 | N/A | 0.1 | 71% (100 cycles)<br>92% (100 cycles) |
| *ACS Sustainable Chem. Eng.* ***12****, 7012–7025 (2024).* | $Li_{2.25}Hf_{0.75}Fe_{0.25}Cl_6$ | $LiCoO_2$ | Li-In alloy | 2.5~4.2 | N/A | 0.1 | 88% (100 cycles) |
| *Nat. Commun.* **14**, 3780 (2023). | $1.5Li_2O$-$HfCl_4$ | NCM83 | Li-In alloy | 2.5~4.3 | 25 | 0.5 | 89.6% (300 cycles) |
| *Nano Lett.* ***22****, 2461–2469 (2022).* | $Li_2HfCl_6$ | NMC532 | Li metal | 2.0~4.2 | 30 | 0.5 | 87% (70 cycles) |